# Deep Learning for Distinguishing Normal versus Abnormal Chest Radiographs and Generalization to Unseen Diseases

Zaid Nabulsi[1], Andrew Sellergren[1†], Shahar Jamshy[1†], Charles Lau[2], Edward Santos[1], Atilla P. Kiraly[1], Wenxing Ye[1], Jie Yang[1], Rory Pilgrim[1], Sahar Kazemzadeh[1], Jin Yu[1], Sreenivasa Raju Kalidindi[3], Mozziyar Etemadi[4], Florencia Garcia-Vicente[4], David Melnick[4], Greg S. Corrado[1], Lily Peng[1], Krish Eswaran[1], Daniel Tse[1*], Neeral Beladia[1], Yun Liu[1], Po-Hsuan Cameron Chen[1*], Shravya Shetty[1*]

[1] Google Health, Google, Palo Alto, CA, USA
[2] Work done at Google Health via Advanced Clinical
[3] Apollo Radiology International, Hyderabad, India
[4] Northwestern Medicine, Chicago, IL, USA
†equal contributions
*corresponding authors: cameronchen@google.com, sshetty@google.com, tsed@google.com

## Abstract

Chest radiography (CXR) is the most widely-used thoracic clinical imaging modality and is crucial for guiding the management of cardiothoracic conditions. The detection of specific CXR findings has been the main focus of several artificial intelligence (AI) systems. However, the wide range of possible CXR abnormalities makes it impractical to build specific systems to detect every possible condition. In this work, we developed and evaluated an AI system to classify CXRs as normal or abnormal. For development, we used a de-identified dataset of 248,445 patients from a multi-city hospital network in India. To assess generalizability, we evaluated our system using 6 international datasets from India, China, and the United States. Of these datasets, 4 focused on diseases that the AI was not trained to detect: 2 datasets with tuberculosis and 2 datasets with coronavirus disease 2019. Our results suggest that the AI system generalizes to new patient populations and abnormalities. In a simulated workflow where the AI system prioritized abnormal cases, the turnaround time for abnormal cases reduced by 7-28%. These results represent an important step towards evaluating whether AI can be safely used to flag cases in a general setting where previously unseen abnormalities exist.

## Introduction

Chest radiography (CXR) is a crucial thoracic imaging modality to detect, diagnose, and guide the management of numerous cardiothoracic conditions. Approximately 837 million CXRs are obtained annually worldwide[1], resulting in a high reviewing burden for radiologists and other healthcare professionals.[2,3] In the United Kingdom, for example, a shortage in the radiology workforce is limiting access to care, increasing wait times, and delaying diagnoses.[4] The need to reduce radiologist workload and improve turnaround time has sparked a surge of interest in developing artificial intelligence (AI)-based tools to interpret CXRs for a broad range of findings.[5–7]

Many algorithms have been shown to detect specific findings, such as pneumonia, pleural effusion, and fracture, with comparable or higher performance than radiologists.[5–10] However, by virtue of being developed to detect specific findings, these algorithms are unlikely to properly report other abnormalities that they were not trained to detect.[11–13] For example, interstitial lung disease may not necessarily trigger a pneumonia detector. If these detectors are indeed highly specific, they can only be used to detect specific diseases, and are not suitable as comprehensive prioritization tools. Moreover, because developing accurate AI algorithms generally requires large labeled datasets, developing algorithms for every potential abnormality that may be encountered in a broad clinical setting is impractical. Therefore, a different problem framing is required for use as an effective prioritization tool: algorithms are needed to distinguish normal versus abnormal CXRs more generally.

A reliable AI system for distinguishing normal CXRs from abnormal ones can contribute to prompt patient workup and management. There are several use cases for such a system. First, in scenarios with a high reviewing burden for radiologists, the AI algorithm could be used to identify cases that are unlikely to contain findings, empowering healthcare professionals to quickly exclude certain differential diagnoses and allowing the diagnostic workup to proceed in other directions without delay. Cases that are likely to contain findings can be also grouped together for prioritized review, reducing the turnaround time. Second, in settings when clinical demand outstrips availability of radiologists (for example, in the midst of a large disease outbreak), such a system might be used as a frontline point-of-care tool for non-radiologists. Importantly, the AI needs to be evaluated on CXRs with "unseen" abnormalities (i.e. those that it had not encountered during development), to validate its robustness towards new diseases or new manifestations of diseases.

In this work, we developed a deep learning system (DLS) that classifies CXRs as normal or abnormal with data from 5 clusters of hospitals from 5 cities in India. We then evaluated the DLS for its generalization[14] to unseen data sources and unseen diseases using 6 independent datasets from India, China, and the United States. These datasets comprise of two broad clinical datasets, two tuberculosis (TB) datasets with microbiologically confirmed positive and negative cases, and two coronavirus disease 2019 (COVID-19) datasets with reverse transcription polymerase chain reaction (RT-PCR)-confirmed positive and negative cases.

## Results

### Dataset curation

Figure 1 shows the overall study design. Our training set consisted of 250,066 CXRs of 213,889 patients from 5 clusters of hospitals from 5 cities in India (Supplementary Table 1, Supplementary Figure 1). In the training set, all known TB cases were excluded and COVID-19 cases were absent. To evaluate the trained DLS, we used 6 datasets with a total of 11,576 CXRs from 11,298 patients (Table 1, Supplementary Figure 1). This includes 2 broad clinical datasets (Dataset 1 [DS-1] and ChestX-ray14 [CXR-14], n=8,557 total cases) with 2,423 abnormal cases, 2 datasets (TB-1 and TB-2, n=595 total cases) with 294 TB-positive cases, and 2 datasets (COV-1 and COV-2, n=2,424 total cases) with 873 COVID-19 positive cases. DS-1, COV-1, and COV-2 were obtained from a mixture of general outpatient and inpatient settings and thus represent a wide spectrum of CXRs seen across different populations. Evaluation on these broad datasets mitigates the risk of selecting only the most obvious cases while excluding more difficult images. CXR-14, TB-1, TB-2 were enriched for rare conditions and were publicly available. Evaluation on these datasets specifically validates the DLS's performance on rarer conditions, and enables benchmarking with other studies using the same data.

To define high-sensitivity and high-specificity operating points for the DLS, we created four small operating point selection datasets for four scenarios: DS-1, CXR-14, TB, and COVID-19; n=200 cases each (see Figure 1B and "Operating point selection datasets" section in Methods). Across these datasets, we collected 48,877 labels from 31 radiologists for either the reference standard or to serve as a comparison for the DLS (see "Labels" section in Methods).

### Classifying CXRs as normal vs abnormal

The DLS was first evaluated for its ability to classify CXRs as normal or abnormal on the test split of DS-1 and an independent test set CXR-14. We obtained the normal and abnormal labels from the majority vote of three radiologists (see "Labels" section in Methods). The percentage of abnormal images were 24% and 71% in DS-1 and CXR-14, respectively (Table 1). The areas under receiver operating characteristic curves (area under ROC, AUC) were 0.87 (95%CI: 0.87-0.88) in DS-1 and 0.94 (95%CI: 0.93-0.96) in CXR-14 (Table 2, Figure 2A). To have a comprehensive understanding of the DLS, we measured sensitivity, specificity, negative predictive value (NPV), positive predictive value (PPV), percentage of predicted positives and the percentage of predicted negatives at a high-sensitivity operating point and a high-specificity operating point ("Evaluation metrics" section in Methods). With the high-sensitivity operating point (see "Operating Point Selection" section in Methods), the DLS predicted 29.9% of DS-1 and 24.0% of CXR-14 as normal, with NPVs of 0.98 and 0.85, respectively (Table 2). With the high-specificity operating point, the DLS predicted 22.2% of DS-1 and 11.7% of CXR-14 as abnormal, with PPVs of 0.68 and 0.99, respectively (Table 2). The NPVs and PPVs across different operating points are plotted in Figure 3.

To put the performance of the DLS in context, two independent board-certified radiologists reviewed both the test split of DS-1 and CXR-14. The radiologists had average NPVs of approximately 0.87 and 0.70 and PPVs of 0.75 and 0.96 on DS-1 and CXR-14, respectively (Table 3). The radiologists' sensitivity and specificity are illustrated on the ROC curves (Figure 2A).

Radiographic findings vary in their difficulty and importance of detection. Thus we next conducted subgroup analyses for each abnormality listed in Supplementary Table 3. The DLS and radiologists' performance for distinguishing normal versus abnormal across all individual findings are shown in Supplementary Figures 2-4 and Supplementary Tables 4 and 5. The DLS showed consistently high NPVs (range: 0.93-1.0) with low variability across all findings in both datasets. The radiologists also showed similar NPVs but with higher variability (range: 0.86-1.0).

## Performance in the setting of unseen diseases

The DLS was next evaluated on two diseases that it had not been trained to detect (TB and COVID-19) across four disease-specific datasets: TB-1, TB-2, COV-1, and COV-2. In these analyses, the DLS was evaluated against the reference standard for each specific disease (TB or COVID, respectively, see "Labels" section in Methods). For TB (where percentage of disease-positive images were 52% and 40% in TB-1 and TB-2; Table 1), the AUCs were 0.95 (95%CIs: 0.93-0.97) in TB-1 and 0.97 (95%CIs: 0.94-0.99) in TB-2 (Table 2, Figure 2B). At the high-sensitivity operating point, the DLS predicted 43.1% of TB-1 and 38.3% of TB-2 as negative, with NPVs of 0.88 and 0.98, respectively (Table 2A). The NPVs and PPVs across different operating points are also plotted in Figure 3. However, CXRs that were labeled (TB) negative could nonetheless contain other abnormalities (see "Labels" section in Methods). Hence PPVs (Table 2A-B) need to be interpreted with the context that low PPVs for identifying TB-positive radiographs as abnormal do not necessarily reflect the PPV for correctly identifying images with other findings in those datasets (see "Distributional shift between datasets" below).

Every image in TB1 and TB2 was also annotated as normal or abnormal by one radiologist from a cohort of 8 consultant radiologists from India. The radiologist NPVs were 0.74 and 0.88 and their PPVs were 0.93 and 0.93 on TB-1 and TB-2, respectively (Table 3 and Figure 2B).

For COVID-19 (where percentage of disease-positive images were 32% and 48% in COV-1 and COV-2; Table 1), the AUCs were 0.68 (95%CIs: 0.66-0.71) in COV-1 and 0.65 (95%CIs: 0.60-0.69) in COV-2 (Table 2, Figure 2A). With a high-sensitivity operating point, the DLS predicts 5.9% of COV-1 and 9.8% of COV-2 as negatives with NPVs of 0.85 and 0.56, respectively (Table 2). The NPVs and PPVs for different operating points are plotted in Figure 3.

Similar to the TB case above, images that were negative for COVID-19 often contained other abnormalities (see "Distributional shift between datasets" section below).

Every image in COV-1 and COV-2 was also reviewed by one radiologist from a cohort of four US board-certified radiologists. The radiologist NPVs were 0.78 and 0.62 and their PPVs were 0.51 and 0.60 on COV-1 and COV-2, respectively (Table 3 and Figure 2C).

Finally, to better understand the potential impact of the algorithm in the setting of imperfect RT-PCR sensitivity, we conducted a subanalysis of COVID-19 cases that had a "false negative" RT-PCR test result on initial testing, defined as a negative RT-PCR test followed by a positive one within five days. In the 21 such cases, the DLS achieved a 95.2% sensitivity, with the CXR taken at the time of the negative test.

## Distributional shifts between datasets

To better understand the data shifts between applications (general clinical setting in DS-1 vs. the enriched CXR-14; the broad clinical settings vs. TB; and the broad clinical settings vs. COVID-19), we next examined the distributions of the DLS predictive scores across all 6 test datasets and their corresponding operating point selection sets (Figure 4, see "Operating Point Selection Datasets" in Methods). We observed similarly peaked DLS prediction score distributions (near 1.0) for positive cases -- whether for general abnormalities, specific conditions, TB, or COVID-19 (see red histograms in Figure 4A-C). However, although the distributions for "negative" cases were mostly similar, they did have a small degree of variability, even among datasets of the same scenario from different sites. For example, comparing TB-1 and TB-2 which have similar CXR findings (TB) but were from two independent sites, negative cases in TB-2 had higher scores than in TB-1. Similarly, comparison between COV-1 and COV-2 also shows slight differences in the scores for negative cases. These observations confirm the existence of data shifts, suggesting that the scenario-specific operating points are essential, and that even having site-specific operating points may further improve the DLS's performance.

Although scores for positive and the negative cases in DS-1, CXR-14, TB-1, and TB-2 were well-separated, there was significant overlap between the distributions of positive and negative cases for the COVID-19 datasets. In fact, further review of the images revealed that 24.9% of negatives in COV-1 and 31.5% of negatives in COV-2 had other CXR findings, and were thus abnormal. A breakdown of the type of finding in these "negatives" is presented in Supplementary Figure 5. Examples of challenging cases of each condition and associated saliency maps highlighting the regions with the greatest influence on DLS predictions are presented in Figure 5.

## Performance of two simulated DLS assisted workflows

To understand how the developed DLS can assist practicing radiologists, we investigated two simulated DLS-based workflows. In the first setup, to assist radiologists in prioritizing review of abnormal cases, the DLS sorted cases by the predicted likelihood of being abnormal (Figure 1D). We measured the differences in expected turnaround time for the abnormal cases with and without DLS prioritization. For simplicity, in this simulation, we assume the same review time for each case, and that the review time per case does not vary based on review order. The DLS-based prioritization reduced the mean turnaround time of abnormal cases by 8-29% for DS-1 and CXR-14, 21-28% for TB-1 and TB-2, and 8-13% for COV-1 and COV-2 (Figure 6). In the second setup, we investigated a simulated sequential reading setup where the DLS identified cases that were unlikely to contain findings, and the radiologist reviewed only the remaining cases (Figure 1D). Though the deprioritized cases could be reviewed at a later time, we computed the effective immediate performance assuming the DLS-negatives were not yet reviewed by radiologists and considered them to be interpreted as "normal" for evaluation purposes. There were minimal performance differences between radiologists and the sequential DLS-radiologists setup, but the effective "urgent" caseload reduced by 25-30% for DS-1 and CXR-14, about 40% for the TB datasets, and about 5-10% for the COVID-19 datasets (Supplementary Table 7).

# Discussion

We have developed and evaluated a DLS for interpreting CXRs as normal or abnormal, instead of detecting individual abnormalities. We further validated that it generalized with acceptable performance using six datasets: two broad clinical datasets (AUC: 0.87 and 0.94), two datasets with one unseen disease (TB; AUC: 0.95 and 0.97), and two datasets with a second unseen disease (COVID-19; AUC: 0.68 and 0.65).

Generalizability to different datasets and patient populations is critical for evaluation of AI systems in medicine. Studies have shown that many factors might lead to challenges of generalization of AI systems to new populations, such as dataset shift and confounders.[15] Furthermore, with CXRs, as with all medical imagery, the number of potential manifestations is unbounded, especially with the emergence of new diseases over time. Understanding model performance on this set of unseen diseases is an imperative step in developing a robust and clinically useful model that can be trusted in real world situations. In this work, we evaluated the DLS's performance on 6 independent test sets consisting of different patient populations, spanning three countries, and with two unseen diseases (TB and COVID-19). The DLS's high sensitivity operating point for ruling out normal CXRs performed on par with board-certified radiologists, with the DLS NPVs of 0.85-0.95 (general abnormalities), 0.88-0.98 (TB), and 0.56-0.85 (COVID-19), comparable to radiologist NPVs of 0.67-0.87 (general abnormalities), 0.74-0.88 (TB), and 0.62-0.78 (COVID-19). These results highlight the DLS's generalizability across real-world dataset shifts, increasing the likelihood of such a system to also generalize to new datasets and new manifestations. The "lower" observed AUCs of the DLS on the COVID-19 datasets were likely caused by our deliberate application of a general abnormality detector to a cohort enriched for patients with a clinical presentation consistent with COVID-19 and thus tested for COVID-19. However, as other acute diseases may share a similar clinical presentation, many cases negative for COVID-19 exhibited abnormal CXR findings that likely triggered the DLS (Figure 5, Supplementary Figure 5). In addition, a substantial number of COVID-19 patients can present with a normal CXR[16], which would also contribute to a lower observed AUC.

The variability in patient population and clinical environment across different datasets also meant that the same operating point was unlikely to be appropriate across all settings. For example, a general outpatient setting is substantially less likely to contain CXR findings compared to a cohort of patients with respiratory symptoms or fevers in the midst of the COVID-19 pandemic. Similarly, datasets that are deliberately enriched for specific conditions (CXR-14 and TB) are skewed and are not representative of a general disease screening population. Thus, we used a small number of cases (n=200) from each setting to determine the operating points specific to that setting. Consistent with this hypothesis, these operating points then generalized well to another dataset, such as from TB-1 to TB-2 and from COV-1 to COV-2. However, further performance improvement is likely possible with site-specific operating point selection sets. We anticipate that this simple operating point selection strategy using a small number of cases may be useful when evaluating an AI system in a new setting, institution, or patient population.

In addition to general performance across the 6 datasets, subgroup analysis of the DLS' performance on each specific abnormal CXR finding of DS-1 and CXR-14 (Supplementary Tables 4 and 5) revealed consistently high NPVs, suggesting that the DLS was not overtly biased towards any particular abnormal finding. In addition, the DLS outperformed radiologists on atelectasis, pleural effusion, cardiomegaly / enlarged cardiac silhouette, and lung nodules - suggesting that the DLS as a prioritization tool could be particularly valuable in emergency medicine where dyspnea, cardiogenic pulmonary edema, and incidental lung cancer detection are commonly encountered. Furthermore, the DLS also outperformed radiologists in settings where an abnormal chest radiographic finding was present but the abnormality was not one of the predefined chest radiographic findings (e.g. perihilar mass) or radiologists agreed on the presence of a finding but disagreed as to its characterization (indicating case ambiguity; see "Other" in Supplementary Tables 4 and 5). This suggests that the DLS may be robust in the setting of chest radiographic findings that are uncommon or difficult to reach consensus on.

To further evaluate the potential utility of our system, we simulated a setup where the DLS prioritizes cases that are likely to contain findings for radiologists' review. Our evaluation suggests a potential reduction in turnaround time for abnormal cases by 7-28%, indicating the DLS's potential to be a powerful first-line prioritization tool. Whether deployed in a relatively healthy outpatient practice or in the midst of an unusually busy inpatient or outpatient setting, such a system could help prioritize abnormal CXRs for expedited radiologist interpretation. In radiology teams where CXR interpretation responsibilities are shared between general and subspecialist (i.e. cardiothoracic) radiologists, such a system could be used to distribute work. For non-radiologist healthcare professionals, a rapid determination regarding the presence or absence of an abnormality on CXR prevents releasing of a patient who needs care and enables alternative diagnostic workup to proceed without delay while the case is pending radiologist review. Finally, a radiologist's productivity might increase by batching negative CXRs for streamlined formal review.

Finally, to facilitate the continued development of AI models for chest radiography, we are releasing our abnormal versus normal labels from 3 radiologists (2430 labels on 810 images) for the publicly-available CXR-14 test set. We believe this will be useful for future work because label quality is of paramount importance for any AI study in healthcare. In CXR-14, the binary abnormal labels were derived through an automated natural language processing (NLP) algorithm on the radiology report.[7] However, editorials have questioned the the quality of labels derived from clinical reports.[17] Hence, in this study we obtained labels from multiple experts to establish the reference standard for evaluation, and a confusion matrix of our majority vote expert labels against the public NLP labels is shown in Supplementary Table 6.

Prior studies have demonstrated an algorithm's potential to differentiate normal and abnormal CXRs.[18–22] Hwang et al. evaluated a commercially available system with comparison to radiology residents.[20] Annarumma et al. further demonstrated the system's utility in a simulated prioritization workflow using held-out data from the same institution as the training dataset.[19] Our study complements prior works by performing extensive evaluations on model generalizability, including generalization to multiple datasets in different continents, different patient populations settings, and with the presence of unseen diseases. In addition, we also obtained radiologist reviews as benchmarks to understand the DLS's performance. Lastly, we presented two simulated workflows; one demonstrated reduced turnaround time for abnormal cases, and the other showed comparable performance while reducing effective caseload.

Our study has several limitations. First, there are a wide range of abnormalities and diseases that were not represented among the CXRs available for this study. Although it's infeasible to exhaustively obtain and annotate datasets for every possible finding, further increasing the conditions and diseases considered in this study could help both in the DLS development and evaluation. Second, we only had labeled data regarding disease-positive and disease-negative for TB and COVID-19. The absence of normal and abnormal labels for the TB and COVID-19 datasets led to added complexity in understanding the performance metrics of PPVs and specificities for these scenarios. Third, to provide a comparison with the DLS, which only had CXRs as input, the radiologists reviewed the cases solely based on CXRs without referencing additional clinical or patient data. In a real clinical setting, this information is generally available, and likely influences a radiologist's decisions. Lastly, the results were based on retrospective data. The utility of the DLS-assisted workflows were based on simulation with many assumptions, such as identical radiologist diagnosis regardless of the review order and identical review time across normal and abnormal cases. Hence, the true effects will need to be determined through future evaluation in a prospective setting.

In conclusion, we have developed and evaluated a clinically relevant artificial intelligence model for chest radiographic interpretation and evaluated its generalizability across a diverse set of images in 6 distinct datasets. These results suggest the potential for the AI system to generalize to new patient populations and unseen diseases. Using the AI system in a simulated workflow to prioritize abnormal cases, the turnaround time for abnormal cases reduced by 7-28%. Lastly, we hope that the performance analyses reported here on the publicly available datasets can serve as a useful resource to facilitate the continued development of clinically useful AI models for CXR interpretation.

# Methods

## Datasets

In this study, we utilized 6 independent datasets for DLS development and evaluation. The DLS was evaluated in two ways: distinguishing normal vs. abnormal cases in a general setting with multiple radiologist-confirmed abnormalities (first 2 datasets), and in the setting of diseases that the DLS was not exposed to during training (TB was excluded from the train set and COVID-19 was not present; last 4 datasets). All data were stored in the Digital Imaging and Communications in Medicine (DICOM) format and de-identified prior to transfer to study investigators. Details regarding these datasets and patient characteristics are summarized in Table 1, Supplementary Table 1, and Supplementary Figure 1. This study using de-identified retrospective data was reviewed by Advarra IRB (Columbia, MD), which determined that it was exempt from further review under 45 CFR 46.

### Train and tune datasets

The first dataset (DS-1) was from five clusters of hospitals across five different cities in India (Bangalore, Bhubaneswar, Chennai, Hyderabad, and New Delhi).[5] DS-1 consisted of images from consecutive inpatient and outpatient encounters between November 2010 and January 2018, and reflected the natural population incidence of the abnormalities in the populations. All TB cases were excluded and COVID-19 cases were not present. In total, DS-1 originally contained 1,052,274 CXRs from 794,501 patients before exclusions (Supplementary Figure 1A). This dataset was randomly split into training, tuning, and testing sets in a 0.775:0.1:0.125 ratio while ensuring that images from the same patient remained in the same split. The split is consistent with our previous study.[5] The DLS was developed solely using the training and tuning splits of DS-1. Because outpatient management is primarily done using posterior-anterior (PA) CXRs, while inpatient management is primarily done on anterior-posterior (AP) CXRs, we emphasized PA CXRs in the tune split to better represent an outpatient use case. Both PA and AP images are used in the test datasets.

### Operating point selection datasets

To select operating points for each of the four scenarios (two general abnormalities, TB, COVID-19), 200 images were randomly selected as the operating point selection sets. For general abnormalities, we selected two independent operating points using 200 randomly sampled images from the DS-1 tune set and 200 randomly sampled images from CXR-14's publicly-specified combined train and tune set[7,23]. For TB, 200 randomly sampled images from TB-1 were used. For COVID-19, 200 randomly sampled images from COV-1 were used. These images were only used to determine an operating point for that scenario, and once used for operating point selection, were excluded from the test set (Supplementary Figure 1).

### Test datasets

Two datasets were used to evaluate the DLS's performance in distinguishing normal and abnormal findings in a general abnormality detection setting. The first dataset contains 7,747 randomly selected PA CXRs from the original test split of the DS-1.[5] These sampled images were expertly labelled as normal or abnormal for the purposes of this study. The second dataset contains 2,000 randomly selected CXRs from the publicly-specified test set (25,596CXRs from 2,797 patients) of CXR-14 from the National Institute of Health.[7,23] From these 2,000 CXRs (also used in prior work[5]), we removed all the patients younger than 18 years of age and all the AP scans (to focus on an outpatient setting, see tune split procedure above), leaving us with 810 images.

To evaluate the DLS performance in unseen diseases, we curated 2 datasets for TB and 2 datasets for COVID-19 (1 CXR per patient, Supplementary Figure 1C-D). For TB, one dataset (TB-1) of 462 PA CXRs with 241 confirmed TB positive CXRs was used, from a hospital in Shenzhen, China. Another dataset (TB-2) of 133 PA CXRs with 53 confirmed TB positive CXRs was used from a hospital in Montgomery, MD, USA.[24–26] Both TB datasets are publicly available. For COVID-19, we used 9,390 CXRs and 5,209 CXRs from all patients who visited two separate hospitals in Chicago in March 2020. Two datasets of 1,819 and 605 AP CXRs (with 583 and 290 CXRs with RT-PCR-confirmed COVID-19 positive diagnoses) were curated from the two hospitals: COV-1, COV-2.

## Labels

### Abnormality labels

For development and evaluation of the DLS, we obtained labels to indicate whether abnormalities were present in each CXR. Each image was annotated as either "normal" or "abnormal", where an "abnormal" scan is defined as a scan containing at least one clinically-significant finding that may warrant further follow-up. For example, degenerative changes and old fractures were not labeled abnormal because no further management is required.

For the train and tune split of DS-1, we obtained the abnormal and normal labels using NLP (regular expressions) on the radiology reports (Supplementary Table 2). For the normal images, radiology report templates were often used, meaning the same report indicating a normal scan was often used for numerous images. We extracted the most commonly used radiology reports, manually confirmed those that indicated normal reports, and obtained all images that used one of these normal template reports. Examples of these radiology reports along with their frequencies are shown in Supplementary Table 2. For the abnormal images, we obtained all images that did not contain keywords indicating the scan is normal in their respective radiology reports.

For the test sets of DS-1 and CXR-14, a group of US board-certified radiologists reviewed the images to provide reference standard labels. For each image in DS-1, three readers were randomly assigned from a cohort of 18 US board-certified radiologists (range of experience 2-24 years in general radiology). For CXR-14, we obtained labels from three US board-certified radiologists (years of experience: 5, 12, and 24). In both cases, the majority vote of the three radiologists was taken to determine the final reference standard label.

For both DS-1 and CXR-14, in addition to the normal versus abnormal label, we also obtained labels for a selected set of findings present in the abnormal images for subgroup analysis (Supplementary Table 3). Note that the lists of findings for DS-1 and CXR-14 differ. For DS-1,we selected a slightly different list of findings to represent conditions that were more clinically reliable, mutually exclusive, and for which the CXR is reasonably sensitive and specific at characterizing (Supplementary Methods and Supplementary Table 3). Similarly to the normal versus abnormal label, the majority vote was taken for each specific finding.  For CXR-14, the differences between the majority voted labels and the publically available labels are shown in a confusion matrix in Supplementary Table 6.

### TB labels

TB positive cases were microbiologically confirmed. The first TB dataset[24] (TB-1) was from Montgomery Country, Maryland, USA, with positive and negative labels from Montgomery County's TB screening results. The second TB dataset[24] (TB-2) was from Shenzhen, China. Positive and negative labels for this dataset came from the TB screening results in the outpatient clinics in Shenzhen No. 3 People's Hospital, Guangdong Medical College, Shenzhen, China.

### COVID-19 labels

For the COVID-19 datasets COV-1 and COV-2, patients with RT-PCR tests and CXRs were included (Supplementary Figure 1). The COVID-19-positive labels were derived from positive RT-PCR tests. In accordance with current Centers for Disease Control and Prevention (CDC) guidelines[27], COVID-19-negative labels consisted of CXRs from patients with at least two consecutive negative RT-PCR tests and no positive test. As false negative rates for RT-PCR have been reported to be ≥20% in symptomatic COVID-19-positive patients, CXRs from patients with only one negative RT-PCR test were excluded.[28]

## Deep learning system development

### Neural network training

We trained a convolutional neural network (CNN) with a single output to distinguish between abnormal and normal CXRs. The CNN uses EfficientNet-B7[29] as its feature extractor, which was pre-trained on 300 million natural images[30]. Since the CNN was pre-trained on three-channel RGB natural images, we tiled the single channel CXR image to three channels for technical compatibility. We trained the CNN using the cross-entropy loss and the momentum optimizer[31] with a constant learning rate of 0.0004 and a momentum value of 0.9. During training, all images were scaled to 600x600 pixels with bilinear interpolation and image pixel values were normalized on a per-image basis to be between 0 and 1. The original bit depth for each image was used (Table 1). For regularization, we applied dropout[32], with a dropout "keep probability" of 0.5. Furthermore, data augmentation techniques were applied to the input images, including horizontal flipping, padding, cropping, and changes in brightness, saturation, hue, and contrast. All hyperparameters were selected based on the empirical performance on the DS-1 tuning set. We developed the network using TensorFlow and used 10 NVIDIA Tesla V100 graphics processing units for training.

### Operating point selection

Given a CXR, the DLS predicts a continuous score between 0 and 1 representing the likelihood of the CXR being abnormal. For making clinical decisions, operating points are needed to threshold the scores and produce binary normal or abnormal categorizations. In this study, we selected two operating points (see "Operating point selection datasets" section above), a high sensitivity operating point (95% sensitivity) and a high specificity operating point (95% specificity) for each scenario: general abnormalities for a general clinical setting in DS-1, general abnormalities for an enriched dataset in CXR-14, TB, and COVID-19.

## Comparison with radiologists

To compare the DLS with radiologists in classifying CXRs as normal versus abnormal, additional radiologists reviewed all test images without referencing additional clinical or patient data. All images in the DS-1 and CXR-14 test set were independently interpreted by two board-certified radiologists (with 2 and 13 years of experience), who classified each CXR as normal or abnormal. These radiologists were independent from the cohort of radiologists who contributed to the reference standard labels.

Each image in TB-1 and TB-2 were reviewed by a random radiologist from a cohort of 8 consultant radiologists in India. Each image was annotated as abnormal or normal. Each image in COV-1 and COV-2 was reviewed by one of four board-certified radiologists (with 2, 5, 13, and 22 years of experience). Similarly, each image was annotated as abnormal or normal.

## Two simulated DLS assisted workflows

We simulated two setups in which the DLS was leveraged to optimize radiologists' workflow (Figure 1D). In the first setup, we randomly sampled 200 CXRs from each of our 6 datasets to simulate a "batch" workload for a radiologist in a busy clinical environment. For these CXRs, we compared the turnaround time for the abnormal CXRs when (1) they were sorted randomly (to simulate a clinical workflow without the DLS) and (2) when the CXRs were sorted in descending order based on the DLS-predicted scores, such that cases with higher scores appeared earlier. We repeated each simulation 1,000 times per dataset to obtain the empirical distribution of turnaround differences.

In the second setup, we analyzed an extreme use case where the DLS identified CXRs that were unlikely to contain findings using a high sensitivity threshold, and the radiologists only reviewed the remaining cases. All cases skipped by radiologists were labeled negative. We compared the sensitivity between this simulated "reduced workload" workflow and a normal workflow in which the radiologists reviewed all cases.

## Evaluation metrics

To evaluate the DLS across different operating points, we calculated the areas under receiver operating characteristic curves (area under ROC, AUC). To evaluate the performance of the DLS in classifying CXRs as normal or abnormal, we measured negative predictive values (NPV), positive predictive values (PPV), sensitivity, specificity, percentage of predicted negatives, and percentage of predicted positives at a high specificity and a high sensitivity operating point chosen for each scenario (see "Operating point selection" in Deep learning system development. For evaluating the DLS for each individual type of finding, we considered a "each abnormality versus normal" setup where negatives consisted of all normal CXRs, and positives consisted of only the CXRs with that particular finding. As such, specificity values were the same across all findings in a given dataset.

We measured the same set of metrics to evaluate the DLS performance with unseen diseases (TB and COVID-19). However, the ground truth here was defined by either the respective TB or COVID-19 tests, and not whether each image contained any abnormal finding. Thus "negative" TB and COVID-19 cases could still contain other abnormalities.

## Statistical analysis

Confidence intervals (CI) for all evaluation metrics were calculated using the non-parametric bootstrap method with n=1,000 permutations at the image level.

To compare the performance of DLS with the radiologists in a DLS-assisted workflow, non-inferiority tests with paired binary data were performed using the Wald test procedure with a 5% margin.[33] To correct for multiple hypothesis testing, we used Bonferroni correction, yielding α=0.003125 (one-sided test with α=0.025 divided by 8 comparisons).[34]

## Saliency map

To provide a visual explanation of how the DLS makes predictions, we utilized gradient-weighted class activation mapping (Grad-CAM)[35] to identify the image regions critical to the model's decision-making process (Figure 5). Because overlaying activation maps on an image obscures the original image, a common Grad-CAM visualization shows two images: the original image, and the image with the overlaid activation maps. Here, to balance brevity and clarity, we present the activation maps as outlines highlighting the regions of interest. The outlines were obtained by taking a horizontal cross-section of the activated maps' three-dimensional contour plot, where the x and y axes represent the spatial location, and the z-axis represents the magnitude of activation.

## Data availability

Many of the datasets used in this study are publicly available. CXR-14 is a public dataset provided by the NIH.[7,23] TB-1 and TB-2 are publicly available.[24] Other than these public datasets, DS-1, COV-1, and COV-2 are owned by their respective institutions and are not publicly available.

## Code availability

The deep learning framework used here (TensorFlow) is available at https://www.tensorflow.org/ and the neural network architecture is available at https://github.com/tensorflow/tpu/tree/master/models/official/efficientnet. The Python libraries used for computation and plotting of the performance metrics (SciPy, NumPy, Lifelines, and Matplotlib) are available from https://www.scipy.org/, http://www.numpy.org/, and https://matplotlib.org/, respectively.

## Acknowledgements

The authors thank the members of the Google Health Radiology and labeling software teams for software infrastructure support, logistical support, and assistance in data labeling. For tuberculosis data collection, thanks go to Sameer Antani, Stefan Jaeger, Sema Candemir, Zhiyun Xue, Alex Karargyris, George R. Thomas, Pu-Xuan Lu, Yi-Xiang Wang, Michael Bonifant, Ellan Kim, Sonia Qasba, and Jonathan Musco. Sincere appreciation also goes to the radiologists who enabled this work with their image interpretation and annotation efforts throughout the study, Jonny Wong for coordinating the imaging annotation work, and David F. Steiner, Kunal Nagpal, and Michael D. Howell for providing feedback on the manuscript.

## Competing interests

This study was funded by Google LLC and/or a subsidiary thereof ('Google'). Z. N., A. S., S. J., E. S., A. P. K., W. Y., J. Y., S. K., J. Y., G. S. C., L. P., K. E., D. T., N. B., Y. L., P.-H. C. C.,and S. S. are employees of Google and own stock as part of the standard compensation package. C. L. is a paid consultant of Google. R. K., M. E., F. G. V., and D. M. received funding from Google to support the research collaboration.

# Tables

**Table 1. Data and patient characteristics of the 6 test datasets.**

| Scenario | Abnormality Detection | | Unseen disease: TB | | Unseen disease: COVID-19 | |
|---|---|---|---|---|---|---|
| | DS-1 | CXR-14 ("ChestX-ray14") | TB-1 | TB-2 | COV-1 | COV-2 |
| Dataset Origin | 5 clusters of hospitals from 5 cities in India | NIH Clinical Center[7] | a hospital in Shenzhen, China | a hospital in Montgomery, MD, USA | a hospital in Illinois, USA | a hospital in Illinois, USA |
| No. Patients | 7,747 | 532 | 462 | 133 | 1,819 | 605 |
| Median Age (IQR) | 48 (38-58) | 49.5 (36-60) | 33 (26-43) | 40 (28-52) | 54 (39-66) | 56 (43-68) |
| No. Female (%) | 2,805 (36.2%) | 375 (46.3%) | 151 (32.7%) | 70 (54.1%) | 950 (47.8%) | 325 (46.3%) |
| Race / ethnicity | N/A | N/A | N/A | N/A | White / Caucasian: 978 (54%) Black / African American: 519 (29%) Asian: 67 (4%) Native Hawaiian / Other Pacific Islander: 5 (0.3%) American Indian / Alaskan Native: 3 (0.2%) Other: 165 (9%) Not Available: 81 (4%) | White / Caucasian: 462 (76%) Black / African American: 58 (10%) Asian: 21 (3%) Native Hawaiian / Other Pacific Islander: 1 (0.2%) American Indian / Alaskan Native: 0 (0%) Other: 53 (9%) Not Available: 10 (2%) |
| No. Images | 7,747 | 810 | 462 | 133 | 1,819 | 605 |
| PA Images | 7,747 | 810 | 462 | 133 | 0 | 0 |
| AP Images | 0 | 0 | 0 | 0 | 1,819 | 605 |
| Reference standard | Normal/abnormal based on majority vote of 3 radiologists | Normal/abnormal based on majority vote of 3 radiologists | TB status based on microbiological confirmation | TB status based on microbiological confirmation | COVID-19 status based on RT-PCR test | COVID-19 status based on RT-PCR test |
| No. abnormal images (%) | 1,845 (23.8%) | 578 (71.4%) | N/A* | N/A* | N/A* | N/A* |
| No. positive images (%, specific disease / finding) | See Supplementary Table 4 | See Supplementary Table 5 | 241 (52.2%, TB) | 53 (39.8%, TB) | 583 (32.1%, COVID-19) | 290 (47.9%, COVID-19) |
| Image properties | | | | | | |
| Width (pixels) | 512–4,400 | 1,143-3,827 | 1,130-3,001 | 4,020-4,892 | 1024-4,200 | 1,024-4,200 |
| Height (pixels) | 512–4,784 | 966–4,715 | 948-3,001 | 4,020-4,892 | 2,014-4,200 | 2,014-4,200 |
| Bit-depth (bits) | 12 | 8 | 8 | 8 | 12 | 12 |

N/A indicates information was not available. * abnormal images in the disease-specific datasets include both those positive for TB or COVID-19, and those with other findings; the numbers of images that contained other findings were not available.

**Table 2. Quantitative evaluation of DLS in distinguishing normal versus abnormal CXRs across 6 datasets. (A)** The DLS's performance with the high-sensitivity operating point. **(B)** The DLS's performance with the high-specificity operating point. The AUC is independent of the operating point and is identical to that in (A).

**A**

| Scenario | Dataset (reference label used for evaluation) | High-sensitivity operating point (optimizes for NPV) | | | | | | AUC (95%CI) |
|---|---|---|---|---|---|---|---|---|
| | | No. predicted negative (%) | NPV (95%CI) | Sensitivity (95%CI) | No. predicted positive (%) | PPV (95%CI) | Specificity (95%CI) | |
| Abnormality detection | DS-1 (normal/ abnormal) | 2313 (29.9%) | 0.98 (0.97-0.99) | 0.98 (0.97-0.98) | 5434 (70.1%) | 0.33 (0.32-0.34) | 0.38 (0.37-0.40) | 0.87 (0.87-0.88) |
| | CXR-14 (normal/ abnormal) | 194 (24.0%) | 0.85 (0.79-0.89) | 0.95 (0.93-0.97) | 616 (76.0%) | 0.89 (0.86-0.91) | 0.71 (0.65-0.76) | 0.94 (0.93-0.96) |
| Unseen disease 1:- TB | TB-1 (TB status) | 199 (43.1%) | 0.88 (0.84-0.93) | 0.90 (0.87-0.94) | 263 (56.9%) | 0.83 0.78-0.87 ) | 0.80 (0.74-0.85) | 0.95 (0.93-0.97) |
| | TB-2 (TB status) | 51 (38.3%) | 0.98 (0.94-1.0) | 0.98 (0.94-1.0) | 82 (61.7%) | 0.63 (0.51-0.73) | 0.63 (0.51-0.73) | 0.97 (0.94-0.99) |
| Unseen disease 2: COVID-19 | COV-1 (COVID-19 status) | 109 (5.9%) | 0.85 (0.78-0.92) | 0.97 (0.96-0.98) | 1710 (94.0%) | 0.33 (0.31-0.35) | 0.08 (0.06-0.09) | 0.68 (0.66-0.71) |
| | COV-2 (COVID-19 status) | 59 (9.8%) | 0.56 (0.43-0.68) | 0.91 (0.87-0.94) | 546 (90.2%) | 0.48 (0.44-0.52) | 0.10 (0.07-0.14) | 0.65 (0.60-0.69) |

**B**

| Scenario | Dataset (reference label used for evaluation) | High-specificity operating point (optimizes for PPV) | | | | | |
|---|---|---|---|---|---|---|---|
| | | No. predicted negative (%) | NPV (95%CI) | Sensitivity (95%CI) | No. predicted positive (%) | PPV (95%CI) | Specificity (95%CI) |
| Abnormality detection | DS-1 (normal/ abnormal) | 6027 (77.8%) | 0.89 (0.88-0.90) | 0.63 (0.61-0.65) | 1720 (22.2%) | 0.68 (0.65-0.70) | 0.91 (0.90-0.91) |
| | CXR-14 (normal/ abnormal) | 715 (88.3%) | 0.32 (0.29-0.36) | 0.16 (0.13-0.20) | 95 (11.7%) | 0.99 (0.96-1.0) | 1.0 (0.99-1.0) |
| Unseen disease1: TB | TB-1 (TB status) | 260 (56.3%) | 0.81 (0.76-0.85) | 0.81 (0.74-0.84) | 202 (43.7%) | 0.95 (0.91-0.98) | 0.95 (0.92-0.98) |
| | TB-2 (TB status) | 80 (60.2%) | 0.94 (0.88-0.99) | 0.91 (0.82-0.98) | 53 (39.8%) | 0.91 (0.81-0.98) | 0.94 (0.88-0.99) |
| Unseen disease 2: COVID-19 | COV-1 (COVID-19 status) | 1558 (85.7%) | 0.72 (0.69-0.74) | 0.23 (0.20-0.27) | 261 (14.3%) | 0.52 (0.46-0.58) | 0.90 (0.88-0.92) |
| | COV-2 (COVID-19 status) | 537(88.8%) | 0.55 (0.51-0.59) | 0.17 (0.12-0.21) | 68 (11.2%) | 0.71 (0.59-0.81) | 0.94 (0.91-0.96) |

**Table 3. Radiologist performance in distinguishing normal and abnormal CXRs across the 6 datasets.**

| Scenario | Dataset (reference label used for evaluation) | Radiologists | | | | | |
|---|---|---|---|---|---|---|---|
| | | No. predicted negative (%) | NPV (95%CI) | Sensitivity (95%CI) | No. predicted positive (%) | PPV (95%CI) | Specificity (95%CI) |
| Abnormality detection | DS-1 (normal/ abnormal) | 6,567 (84.8%) | 0.86 (0.85-0.86) | 0.48 (0.46-0.51) | 1,180 (15.2%) | 0.76 (0.74-0.78) | 0.95 (0.95-0.96) |
| | | 6,380 (82.4%) | 0.87 (0.86-0.88) | 0.54 (0.52-0.57) | 1,367 (17.6%) | 0.74 (0.71-0.76) | 0.94 (0.93-0.94) |
| | CXR-14 (normal/ abnormal) | 284 (35.1%) | 0.73 (0.67-0.77) | 0.87 (0.84-0.89) | 526 (64.9%) | 0.95 (0.93-0.97) | 0.89 (0.85-0.93) |
| | | 325 (40.1%) | 0.67 (0.62-0.72) | 0.81 (0.78-0.84) | 485 (59.9%) | 0.97 (0.96-0.99) | 0.94 (0.91-0.97) |
| Unseen disease: TB | TB-1 (TB status) | 282 (61.0%) | 0.74 (0.69-0.80) | 0.70 (0.65-0.76) | 180 (39.0%) | 0.93 (0.89-0.97) | 0.95 (0.91-0.97) |
| | TB-2 (TB status) | 88 (66.2%) | 0.88 (0.81-0.94) | 0.79 (0.68-0.90) | 45 (33.8%) | 0.93 (0.85-1.0) | 0.96 (0.92-1.0) |
| Unseen disease: COVID-19 | COV-1 (COVID-19 status) | 1,194 (65.6%) | 0.78 (0.76-0.80) | 0.55 (0.51-0.59) | 625 (34.4%) | 0.51 (0.47-0.54) | 0.75 (0.73-0.77) |
| | COV-2 (COVID-19 status) | 352 (58.2%) | 0.62 (0.57-0.66) | 0.53 (0.48-0.59) | 253 (41.8%) | 0.60 (0.55-0.66) | 0.68 (0.64-0.74) |

# Figures

**A. Training and tuning**

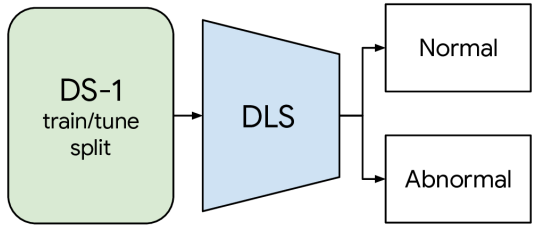

**B. Operating points selection**

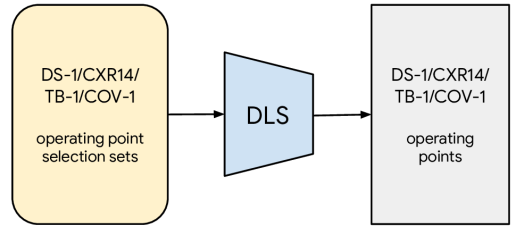

**C. Deep learning system (DLS) and radiologists evaluation**

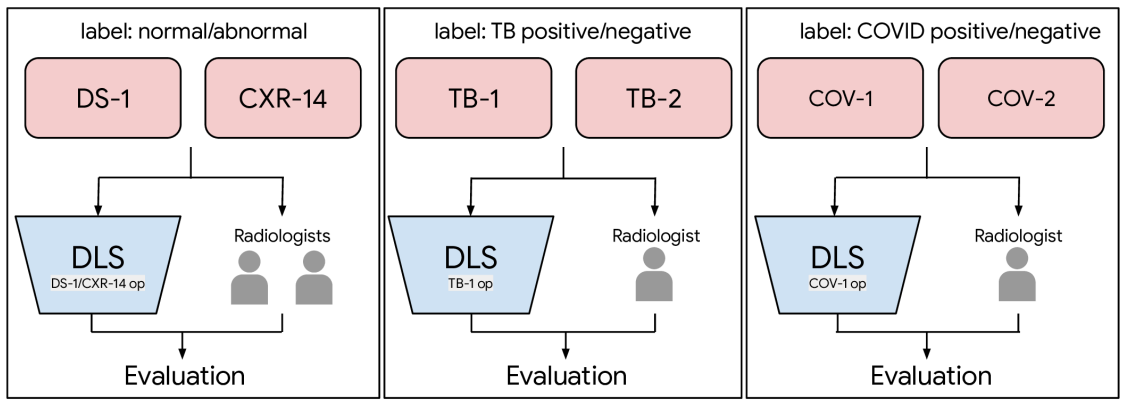

**D. DLS + Radiologist Simulated Workflows**

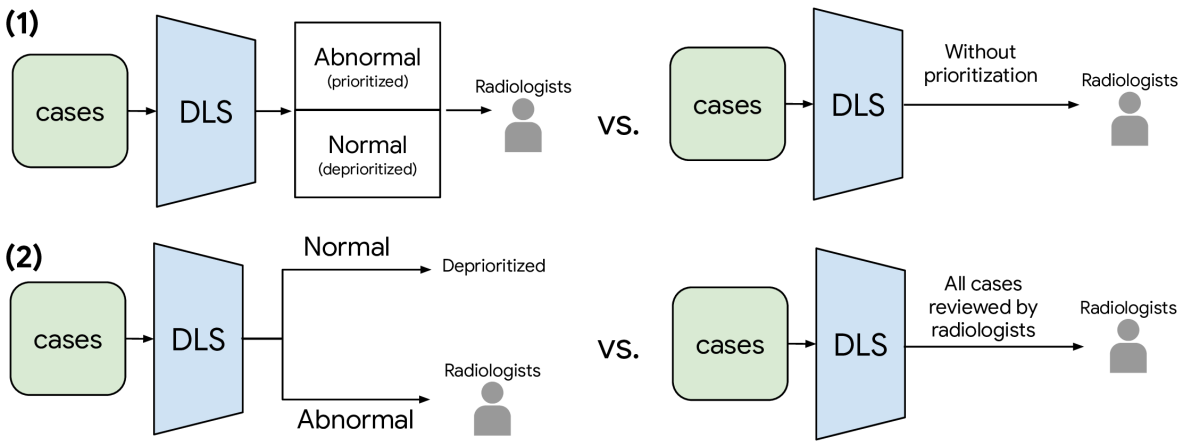

**Figure 1**. **Schematic of the study design, including (A) training and tuning, (B) operating points selection, (C) evaluation on the deep learning system and radiologists, and (D) two simulated DLS-assisted workflows.** DS-1, CXR-14, TB-1, TB-2, COV-1, COV-2 are abbreviations of the datasets, please see Table 1 and Supplementary Table 1 for details.

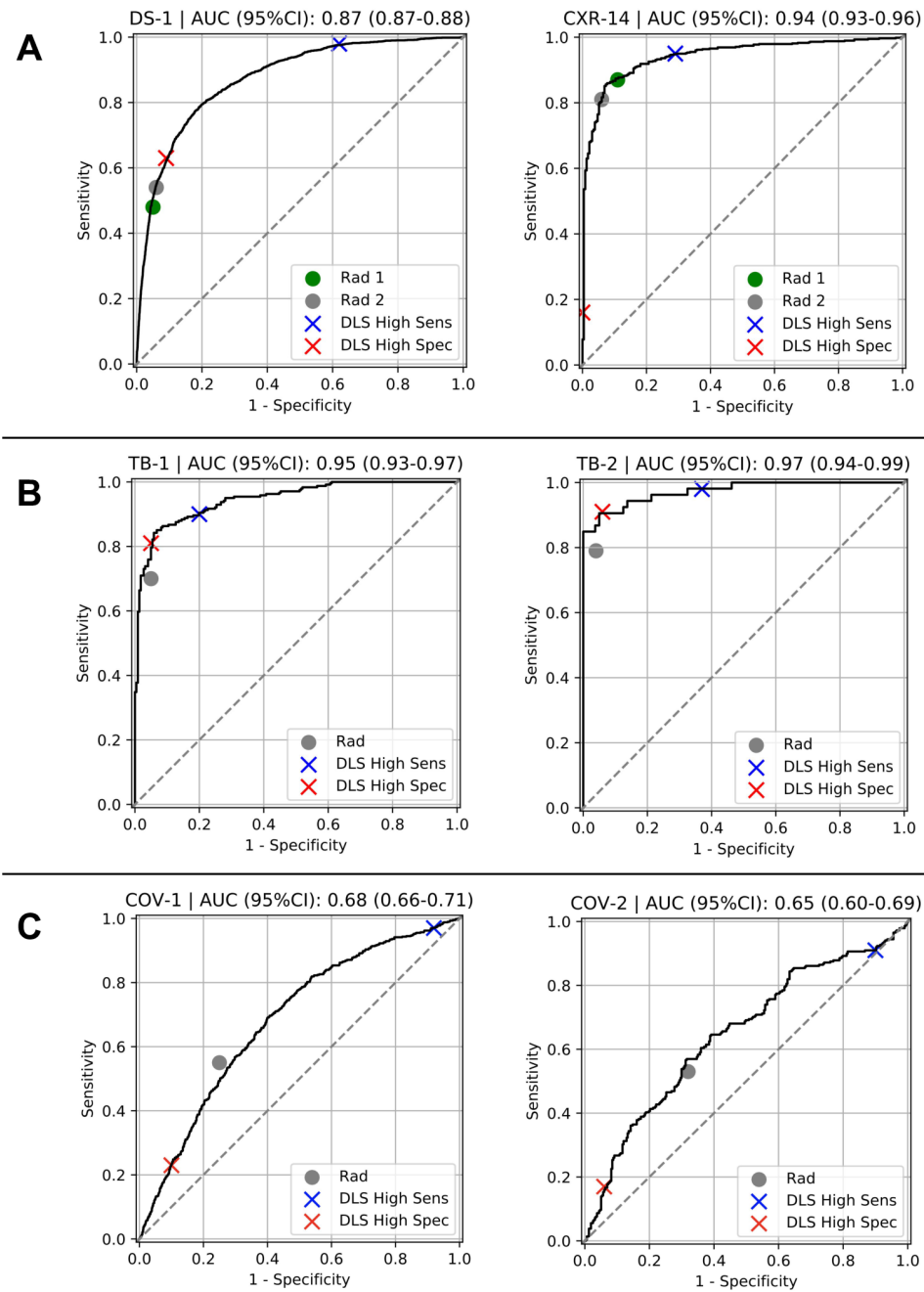

**Figure 2. Receiver operating characteristic (ROC) curves for the DLS in distinguishing normal and abnormal CXRs across 6 different datasets.** Positive CXRs in DS-1 and CXR-14 contain a mix of multiple labeled abnormalities (Supplementary Table 3). Positive CXRs in the two TB datasets are from patients with tuberculosis. Positive CXRs in the two COVID-19 datasets are from patients with reverse transcription polymerase chain reaction (RT-PCR)-verified COVID-19. Radiologists' performances in distinguishing the test cases as normal or abnormal are also highlighted in the figures.

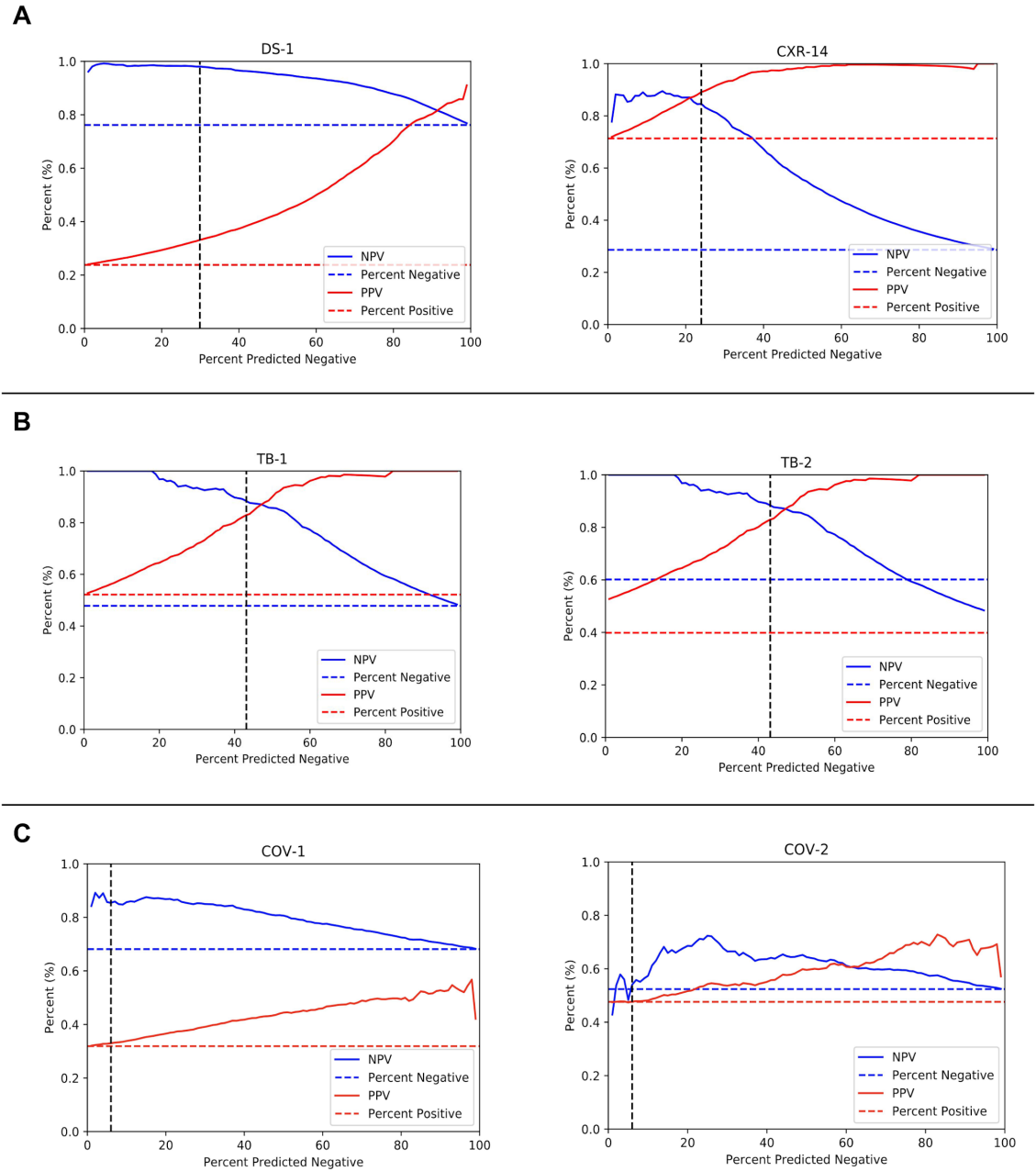

**Figure 3. Positive predictive values (PPV) and negative predictive values (NPV) of the DLS across 6 datasets. (A) General abnormalities: DS-1 and CXR-14 datasets. (B) TB: TB-1 and TB-2. (C) COVID-19: COV-1 and COV-2.** The horizontal dotted lines represent the prevalence of positive examples (red) and negative examples (blue), which also correspond to random predictions' PPV and NPV, respectively. The DLS's NPV converges to the prevalence of negative examples when all examples are predicted as negative, and the DLS's PPV converges to the prevalence of positive examples when all examples are predicted as positive. The vertical, dotted black lines highlight the selected operating point at 95% sensitivity on the operating point selection sets for each scenario.

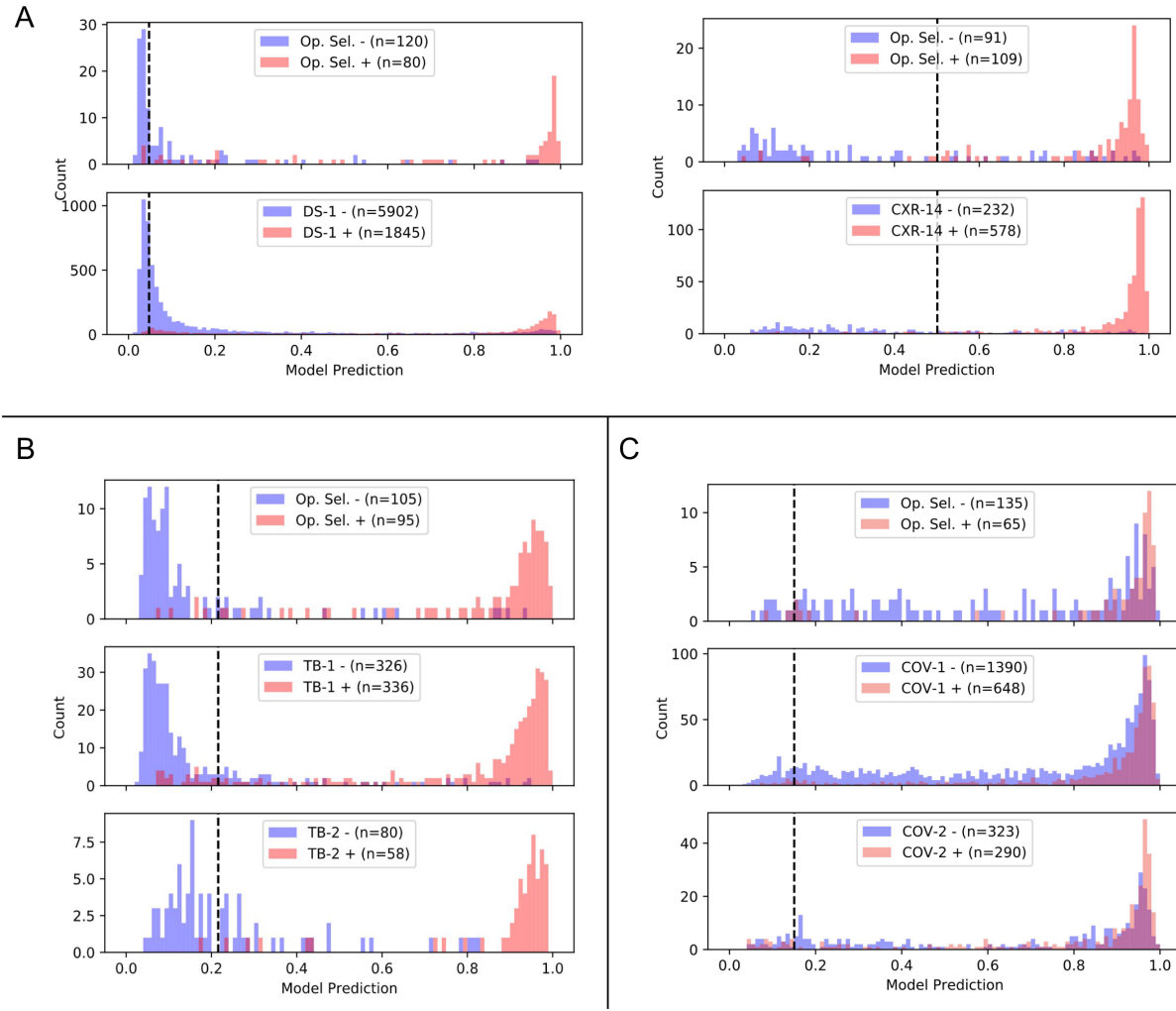

**Figure 4. Histogram for the distribution of DLS predicted scores across 6 datasets and their corresponding operating point selection sets: (A) DS-1 and CXR-14, (B) TB-1 and TB-2, and (C) COV-1 and COV-2.** Curation of the operating point selection (Op. Sel.) datasets is described in "Operating point selection datasets" in Methods. Positive and negative examples are visualized separately in red and blue, respectively. The vertical lines (black) highlight the selected high-sensitivity operating point for each scenario.

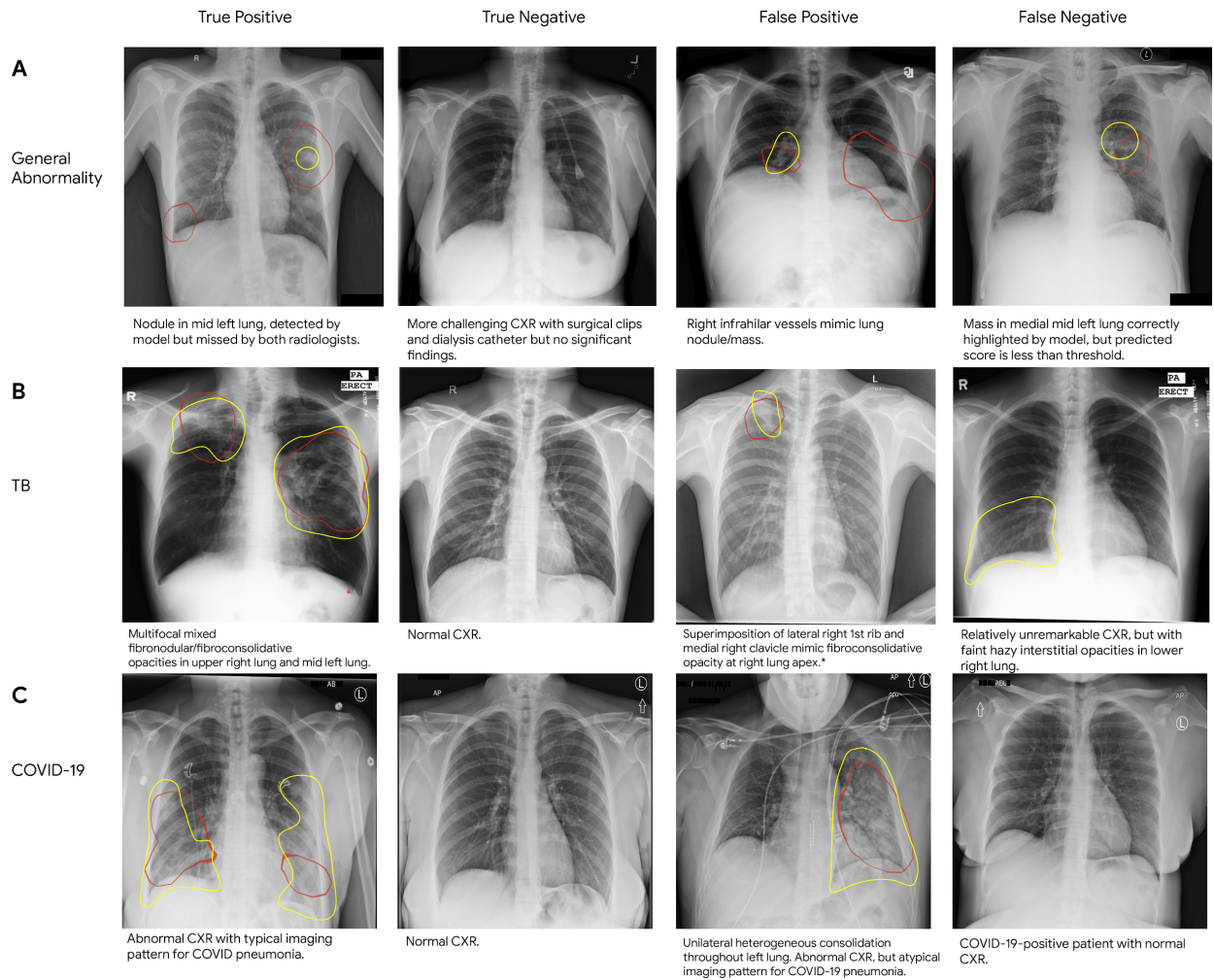

**Figure 5. Sample CXRs of true and false positives, and true and false negatives for (A) general abnormalities, (B) TB, and (C) COVID-19.** Each image has the saliency presented as red outlines that indicate the areas the DLS is focusing on for identifying abnormalities, and yellow outlines representing regions of interest indicated by radiologists. Text descriptions for each CXR are below the respective image. Note that the general abnormality false negative example is shown with abnormal saliency maps. However, the DLS predictive score on the case was lower than the selected threshold; hence the image was classified as "normal". *Note that the TB false positive image was saved in the system with inverted colors, and presented to the model that way. Colors have been uninverted for visualization purposes.

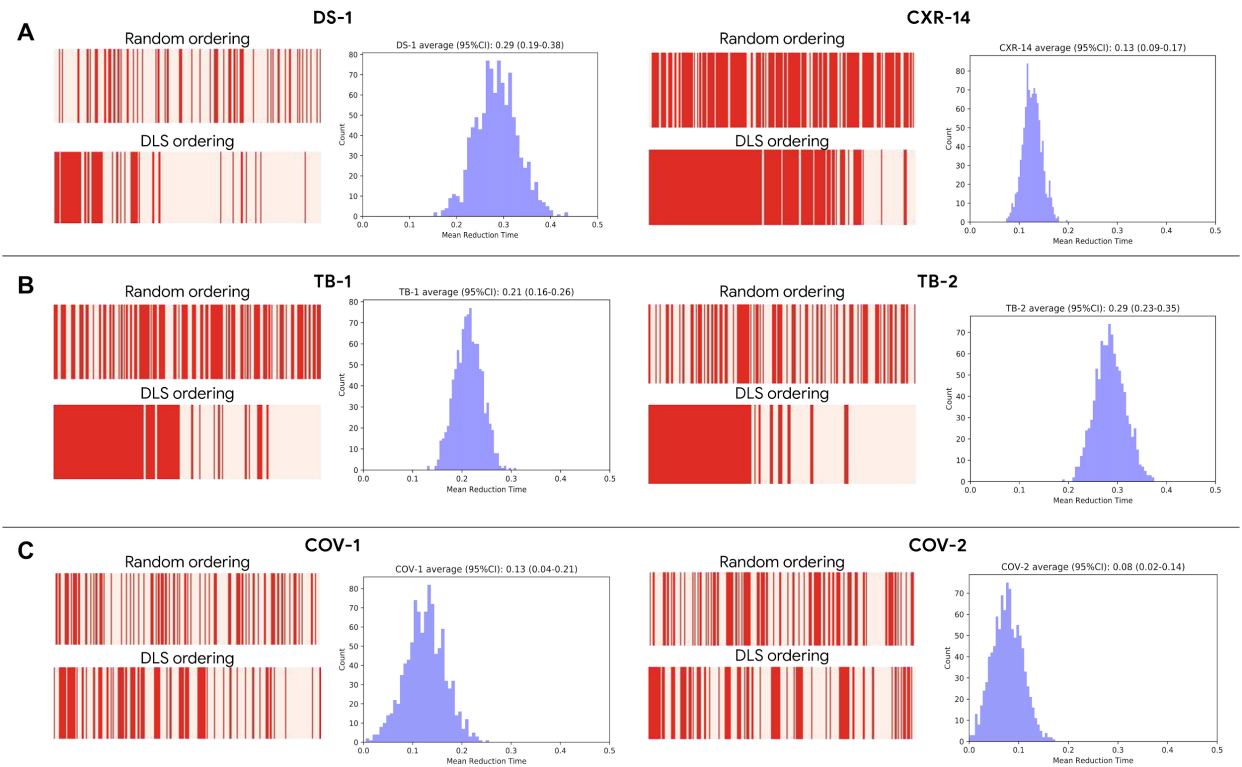

**Figure 6. Impact of a simulated DLS-based prioritization in comparison with random review order for (A) general abnormalities, (B) TB, and (C) COVID-19.** The red bars indicate sequences of abnormal CXRs in red and normal CXRs in pink; a greater density of red towards the left indicates abnormal CXRs are reviewed sooner than normal ones. The histograms indicate the average improvement in turnaround time.

# Supplementary information

## Supplementary Methods

**List of specific findings for DS-1**

We modified the list of findings from CXR-14 to include conditions that were more likely to be clinically actionable, mutually exclusive, and for which CXR is reasonably sensitive and specific for characterizing (Supplementary Table 3). For example, findings in CXR-14 such as "emphysema" (for which CXR lacks both sensitivity and specificity) and "infiltration" (an ambiguous term that overlaps other CXR-14 findings such as "pneumonia" and "atelectasis") were replaced by more specific terms. On the other hand, clinically relevant and distinct findings commonly encountered on CXR were also introduced (e.g. "hilar enlargement", "acute fracture") or augmented (e.g. "abnormal mediastinal mass/widening" rather than "hiatal hernia"). Our choice of findings for the DS-1 dataset also recognized inherent limitations of CXR for reliably distinguishing between some conditions; hence "focal/multifocal lung opacity" was adopted as a single finding, rather than distinct findings for "consolidation", "atelectasis", and "fibroconsolidative opacity".

# Supplementary Figures

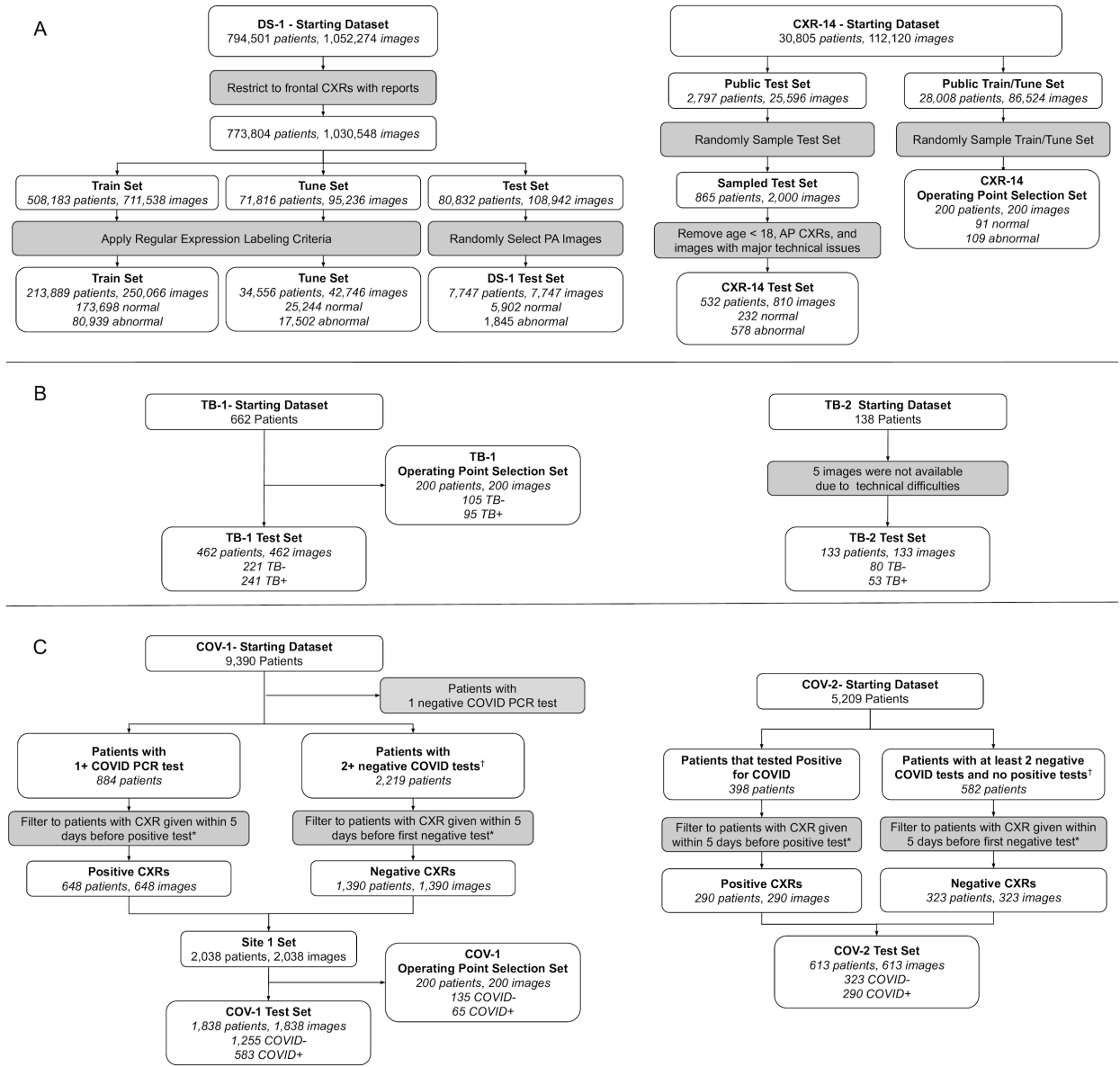

**Supplementary Figure 1. The STARD diagrams with inclusion and exclusion criteria for the 6 datasets.** *For COVID-19, the first CXR during the patient's hospital encounter was selected. †Negative tests had to be administered at least 12 hours apart.

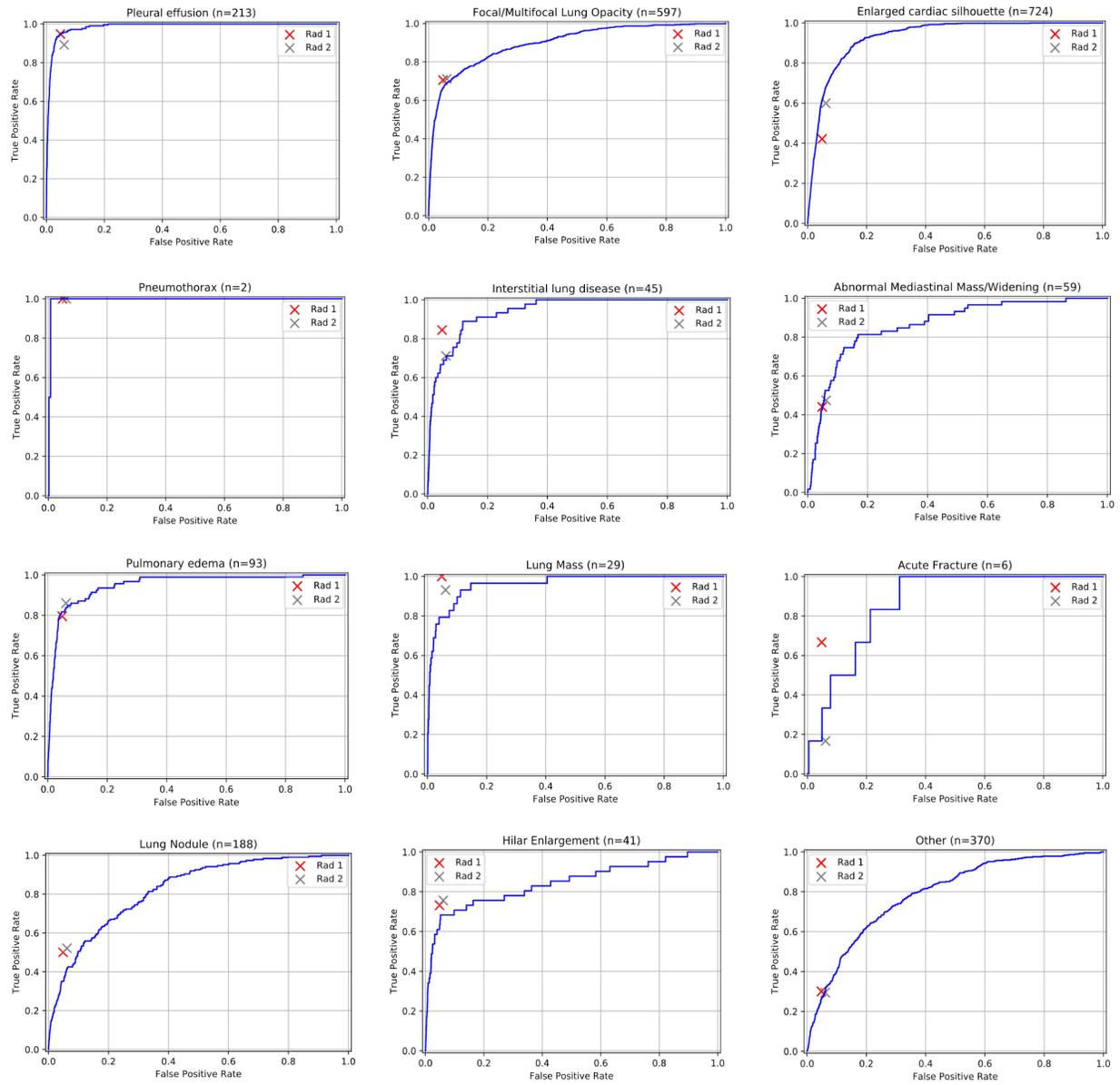

**Supplementary Figure 2. The ROC curves of the DLS in detecting specific findings in DS-1.**

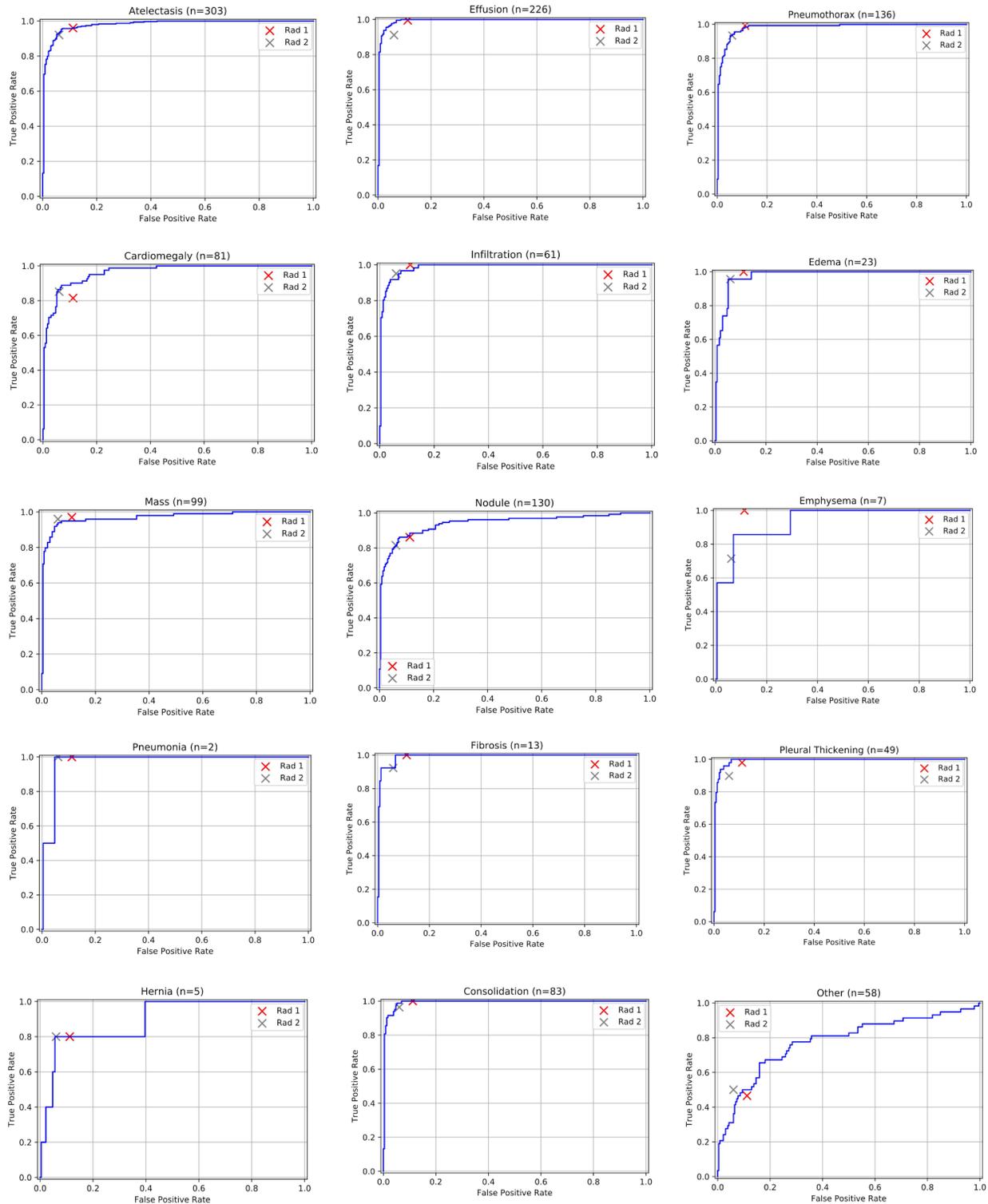

**Supplementary Figure 3. The ROC curves of the DLS in detecting specific findings in CXR-14.**

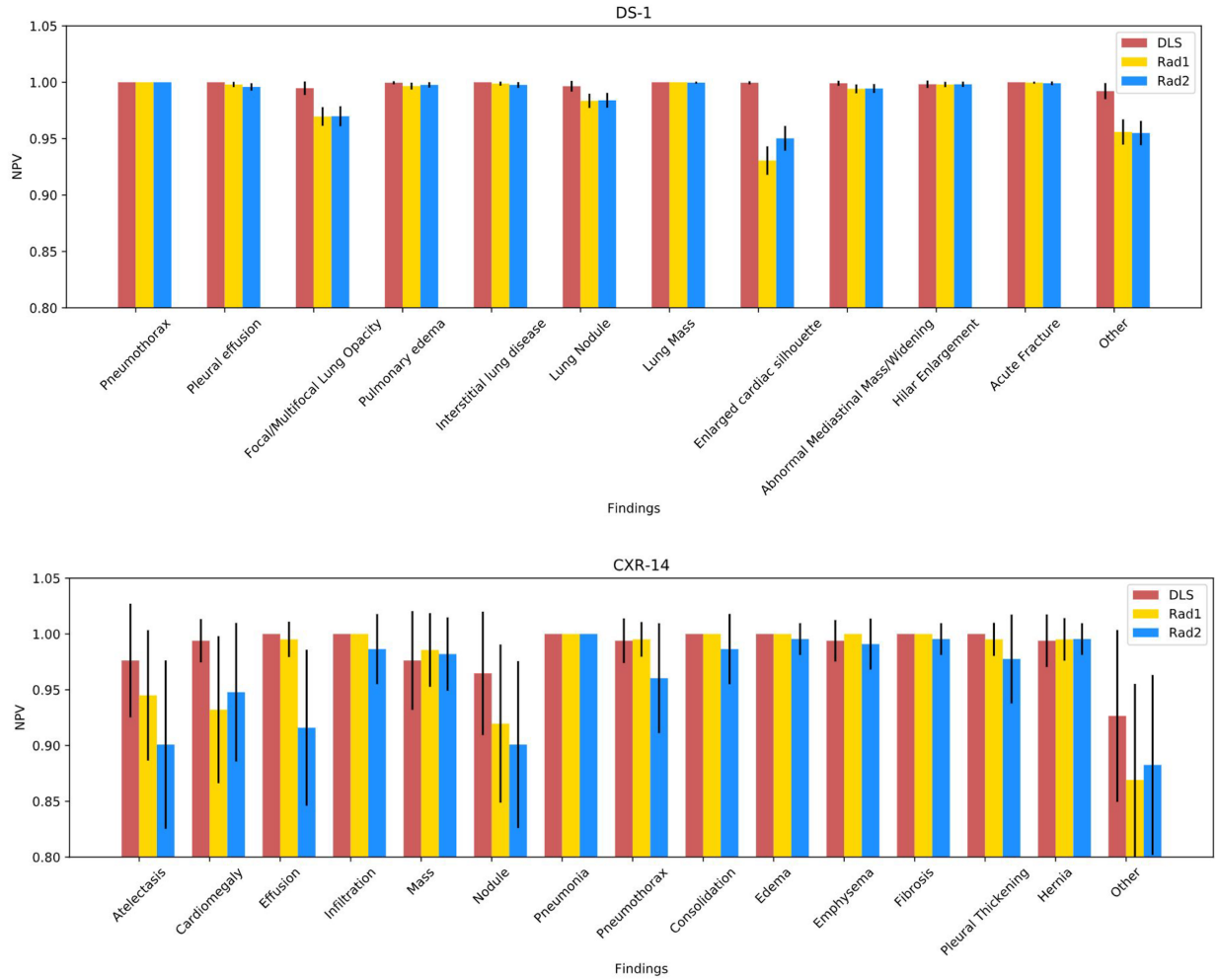

**Supplementary Figure 4. Comparison of NPVs between the DLS and the radiologists across specific findings in DS-1 and CXR-14.**

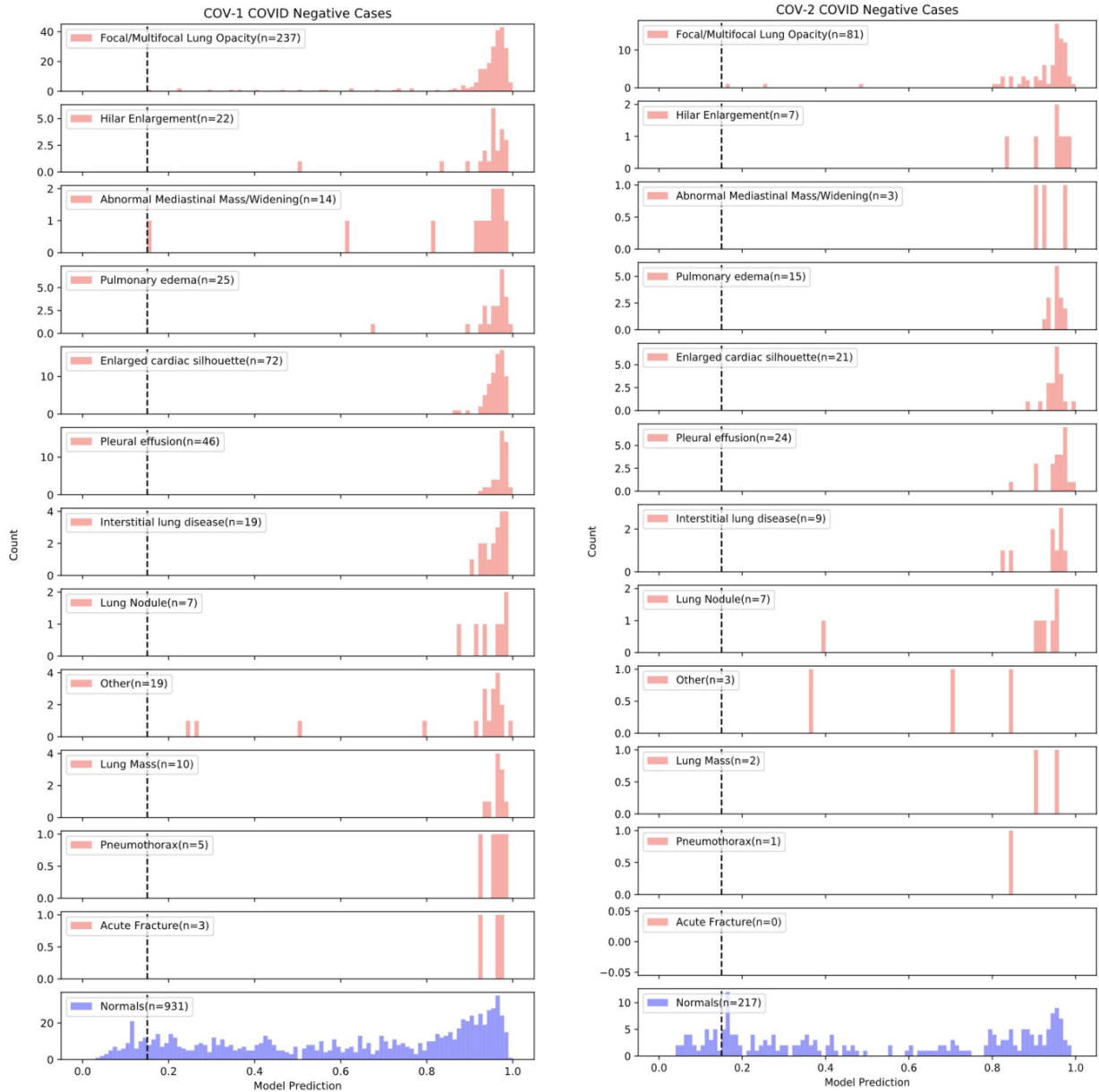

**Supplementary Figure 5. Histograms for the distributions of DLS-predicted scores for specific findings in COVID-19-negative cases in COV-1 and COV-2.** The findings were indicated by U.S. board-certified radiologists and were not mutually exclusive as a single case may have had multiple findings.

## Supplementary Tables

**Supplementary Table 1. Data and patient characteristics for the train set, tune set, and operating point selection sets.** *Information not available; for TB and COVID-19 datasets, cases without the disease may still have other abnormal findings.

| Datasets | DS-1 train split | DS-1 tune split | DS-1 operating point selection set | CXR-14 operating point selection set | TB-1 operating point selection set | COV-1 operating point selection set |
|---|---|---|---|---|---|---|
| Dataset Origin | 5 clusters of hospitals from 5 cities in India | 5 clusters of hospitals from 5 cities in India | 5 clusters of hospitals from 5 cities in India | NIH Clinical Center | a hospital in Shenzhen, China | a hospital in Illinois, USA |
| No. Patients | 213,889 | 34,556 | 200 | 181 | 200 | 200 |
| Median Age (IQR) | 48 (37-58) | 49 (39-59) | 49 (39-59) | 49 (37-59) | 32 (25-42) | 56 (38-67) |
| No. Female (%) | 91,654 (36.7%) | 16092 (37.6%) | 76 (38.0%) | 77 (38.5%) | 62 (31%) | 99 (50.5%) |
| Race / ethnicity | N/A | N/A | N/A | N/A | N/A | White / Caucasian: 114 (57%) Black / African American: 55 (28%) Asian: 5 (3%) Native Hawaiian / Other Pacific Islander: 0 (0%) American Indian / Alaskan Native: 0 (0%) Other: 20 (10%) Not Available: 6 (3%) |
| No. Images | 250,066 | 42,746 | 200 | 200 | 200 | 200 |
| PA Images | 202,681 | 42,746 | 200 | 124 | 0 | 200 |
| AP Images | 47,385 | 0 | 0 | 76 | 200 | 0 |
| Number of abnormal images | 80,939 | 17,502 | 80 | 109 | N/A* | N/A* |
| Disease/finding no. of positive images | N/A | N/A | N/A | N/A | TB+:95 | COVID+: 65 |

N/A indicates Information was not available. *For the TB and COVID-19 datasets, cases without the disease may still have other abnormal findings.

**Supplementary Table 2. Radiology report pattern matching. (A)** Normal radiology report templates. The five-most used radiology reports to indicate that a scan is normal, along with the number of occurrences in the DS-1 train set. To obtain normal examples in our train set, we extracted all radiology reports that were used in a substantial number of cases (at least 50 occurrences) in the set and manually verified each report to ensure that it represented a "normal" scan. We then used all the images that had those radiology reports as "normal" (negative) examples in our train set. **(B)** regular expressions (natural language processing) used for identifying abnormal cases.

**A**

| Unparsed Radiology Report Template | Number of Occurrences in DS-1 Train |
|---|---|
| Provisional Diagnosis/Clinical Data : NILReport:: Lung fields are clear. Cardio thoracic ratio is normal. Apices, costo and cardiophrenic angles are free. Cardio vascular shadow and hila show no abnormal feature. Bony thorax shows no significant abnormality. Domes of diaphragm are well delineated.Impression –Normal Study | 60,935 |
| Report:: Lung fields are clear. Cardio thoracic ratio is normal. Apices, costo and cardiophrenic angles are free. Cardio vascular shadow and hila show no abnormal feature. Bony thorax shows no significant abnormality. Domes of diaphragm are well delineated.Impression –Normal Study | 56,114 |
| Observation Both lung fields clear. Both hila are normal in size and position. Costophrenic and cardiophrenic angles are clear. Cardiac size and contour are within normal limits. Rib cage is normal. Soft tissues are normal.Impression Normal study. | 12,537 |
| Report:: Lung fields are clear. Cardio thoracic ratio is normal. Apices, costo and cardiophrenic angles are free. Cardio vascular shadow and hila show no abnormal feature. Bony thorax shows no significant abnormality. Domes of diaphragm are well delineated.Impression 1. Normal Study. | 9,568 |
| Report :: Lung fields are clear. Cardio thoracic ratio is normal. Apices, costo and cardiophrenic angles are free. Cardio vascular shadow and hila show no abnormal feature. Bony thorax shows no significant abnormality. Domes of diaphragm are well delineated. IMPRESSION NORMAL STUDY' | 3,829 |

**B**

| Regular expression | Number of cases that do not contain the regular expression in DS-1 train |
|---|---|
| "Normal\s+study" | 39,628 |
| "Impression.+no significant abnormality" | 45,387 |

**Supplementary Table 3. List of findings in DS-1 and ChestX-ray14.**

| DS-1 Findings | ChestX-ray14 Findings | Comments |
|---|---|---|
| Pneumothorax | Pneumothorax | — |
| Pleural effusion | Effusion | — |
| — | Pleural thickening | CXR is generally nonspecific for pleural thickening, an entity that is often not clinically actionable. |
| — | Emphysema | CXR is not sensitive and not specific for emphysema. |
| — | Infiltration | Deprecated term whose use is discouraged by most subspecialists in thoracic imaging. Definition may vary by individual, and may be taken to imply any/some/all of the following: consolidation, fibrosis, atelectasis, pulmonary edema. |
| Focal/multifocal lung opacity | Consolidation | Consolidation refers to a homogeneous lung opacity that obscures vessel and airway wall margins. Common causes of consolidation include pneumonia, pulmonary edema, and alveolar hemorrhage - entities that are often difficult to distinguish from each other on CXR. As atelectasis and lung fibrosis may often be difficult to distinguish from consolidation on CXR, aggregating these entities may strike a reasonable balance between achieving inter-observer agreement and acknowledging the limitations of CXR. |
| Focal/multifocal lung opacity | Pneumonia | |
| Focal/multifocal lung opacity | Fibrosis | |
| Focal/multifocal lung opacity | Atelectasis | |
| Pulmonary edema | Edema | — |
| Interstitial lung disease | — | Interstitial lung disease is a pathologic entity involving the supporting framework of the lung, and may occur in the setting of occupational inhalation exposures or in older individuals with unexplained chronic dyspnea. |
| Lung nodule | Nodule | — |
| Lung mass | Mass | — |
| Enlarged cardiac silhouette | Cardiomegaly | Enlarged cardiac silhouette can on occasion be caused by pericardial effusion rather than cardiomegaly. As CXR is relatively nonspecific for cardiomegaly vs. pericardial effusion, "enlarged cardiac silhouette" may acknowledge limitations of CXR better. |
| Abnormal mediastinal mass/widening | Hernia | A hiatal hernia is one of many causes of abnormal mediastinal mass/widening, which can range from clinically insignificant to life-threatening. |
| Hilar enlargement | — | Hilar enlargement is often an actionable imaging finding, which may be caused by malignant, infectious, or inflammatory lymphadenopathy or pulmonary arterial hypertension. |
| Acute fracture | — | Acute fractures are common findings encountered in the emergency/trauma setting and may be associated with other actionable secondary diagnoses (e.g. pneumothorax or hemothorax) or portend more serious underlying injuries. |

# Supplementary Table 4. Performance of DLS on specific findings on DS-1.

| Findings | # positives | High-sensitivity operating point | | | High-specificity operating point | | | AUC [95% CI] | Radiologist 1 | Radiologist 2 |
|---|---|---|---|---|---|---|---|---|---|---|
| | | % predicted negative | NPV | Sensitivity | % predicted positive | PPV | Specificity | | | |
| Pneumothorax | 2 | 38.4% | 1 | 1 | 8.7% | 0.004 | 0.91 | 1.0 [0.99, 1.0] | NPV: 1.0<br>Sens: 1.0<br>Spec: 0.95<br>PPV: 0.007 | NPV: 1.0<br>Sens: 1.0<br>Spec: 0.94<br>PPV: 0.005 |
| Pleural Effusion | 213 | 37.1% | 1 | 1 | 11.8% | 0.27 | 0.91 | 0.99 [0.98, 0.99] | NPV: 0.86<br>Sens: 0.48<br>Spec: 0.95<br>PPV: 0.76 | NPV: 1.0<br>Sens: 0.89<br>Spec: 0.94<br>PPV: 0.34 |
| Focal or Multifocal Lung Opacity | 597 | 35.1% | 0.99 | 0.98 | 14.6% | 0.44 | 0.91 | 0.90 [0.89, 0.91] | NPV: 0.97<br>Sens: 0.71<br>Spec: 0.95<br>PPV: 0.60 | NPV: 0.97<br>Sens: 0.71<br>Spec: 0.94<br>PPV: 0.54 |
| Pulmonary Edema | 93 | 37.8% | 1.0 | 0.99 | 9.92% | 0.13 | 0.91 | 0.95 [0.92, 0.97] | NPV: 1.0<br>Sens: 0.8<br>Spec: 0.95<br>PPV: 0.21 | NPV: 1.0<br>Sens: 0.86<br>Spec: 0.94<br>PPV: 0.18 |
| Interstitial Lung Disease | 45 | 38.1% | 1.0 | 1.0 | 9.23% | 0.06 | 0.91 | 0.94 [0.91, 0.96] | NPV: 1.0<br>Sens: 0.84<br>Spec: 0.95<br>PPV: 0.12 | NPV: 1.0<br>Sens: 0.71<br>Spec: 0.94<br>PPV: 0.08 |
| Lung Nodule | 188 | 37.3% | 1.0 | 0.96 | 8.84% | 0.14 | 0.91 | 0.82 [0.80, 0.85] | NPV: 0.98<br>Sens: 0.50<br>Spec: 0.95<br>PPV: 0.25 | NPV: 0.98<br>Sens: 0.52<br>Spec: 0.94<br>PPV: 0.21 |
| Lung Mass | 29 | 38.2% | 1.0 | 1.0 | 9.09% | 0.04 | 0.91 | 0.96 [0.93, 0.98] | NPV: 1.0<br>Sens: 1.0<br>Spec: 0.95<br>PPV: 0.09 | NPV: 1.0<br>Sens: 0.93<br>Spec: 0.94<br>PPV: 0.07 |
| Enlarged Cardiac Silhouette | 724 | 34.2% | 1.0 | 1.0 | 16.0% | 0.50 | 0.91 | 0.93 [0.92, 0.94] | NPV: 0.93<br>Sens: 0.42<br>Spec: 0.95<br>PPV: 0.52 | NPV: 0.95<br>Sens: 0.60<br>Spec: 0.94<br>PPV: 0.54 |
| Abnormal Mediastinal Mass/Widening | 59 | 38.1% | 1.0 | 0.97 | 9.23% | 0.06 | 0.91 | 0.87 [0.82, 0.91] | NPV: 0.99<br>Sens: 0.43<br>Spec: 0.95<br>PPV: 0.08 | NPV: 0.99<br>Sens: 0.46<br>Spec: 0.94<br>PPV: 0.07 |
| Hilar Enlargement | 41 | 38.1% | 1.0 | 0.90 | 9.14% | 0.05 | 0.91 | 0.84 [0.76, 0.92] | NPV: 1.0<br>Sens: 0.73<br>Spec: 0.95<br>PPV: 0.09 | NPV: 1.0<br>Sens: 0.76<br>Spec: 0.94<br>PPV: 0.08 |
| Acute Fracture | 6 | 38.4% | 1.0 | 1.0 | 8.77% | 0.005 | 0.91 | 0.86 [0.76, 0.96] | NPV: 1.0<br>Sens: 0.67<br>Spec: 0.95<br>PPV: 0.01 | NPV: 1.0<br>Sens: 0.17<br>Spec: 0.94<br>PPV: 0.002 |
| Other (i.e., not a finding listed above) | 377 | 36.4% | 0.99 | 0.95 | 10.4% | 0.20 | 0.91 | 0.79 [0.77, 0.81] | NPV: 0.96<br>Sens: 0.31<br>Spec: 0.95<br>PPV: 0.29 | NPV: 0.95<br>Sens: 0.29<br>Spec: 0.94<br>PPV: 0.23 |

**Supplementary Table 5. Performance of DLS on specific findings on CXR-14.**

| Findings | # positives | High-sensitivity operating point | | | High-specificity operating point | | | AUC [95% CI] | Radiologist 1 | Radiologist 2 |
|---|---|---|---|---|---|---|---|---|---|---|
| | | % predicted negative | NPV | Sensitivity | % predicted positive | PPV | Specificity | | | |
| Atelectasis | 303 | 31.4% | 0.98 | 0.99 | 14.4% | 0.99 | 1.0 | 0.98 [0.97, 0.99] | NPV: 0.94<br>Sens: 0.96<br>Spec: 0.89<br>PPV: 0.92 | NPV: 0.90<br>Sens: 0.92<br>Spec: 0.94<br>PPV: 0.95 |
| Cardiomegaly | 81 | 52.8% | 0.99 | 0.99 | 5.75% | 0.94 | 1.0 | 0.96 [0.94, 0.98] | NPV: 0.93<br>Sens: 0.81<br>Spec: 0.89<br>PPV: 0.72 | NPV: 0.95<br>Sens: 0.85<br>Spec: 0.94<br>PPV: 0.83 |
| Effusion | 226 | 35.8% | 1.0 | 1.0 | 15.7% | 0.99 | 1.0 | 0.99 [0.97, 1.0] | NPV: 1.0<br>Sens: 1.0<br>Spec: 0.89<br>PPV: 0.90 | NPV: 0.92<br>Sens: 0.91<br>Spec: 0.94<br>PPV: 0.94 |
| Infiltration | 61 | 56.0% | 1.0 | 1.0 | 4.10% | 0.92 | 1.0 | 0.99 [0.97, 1.0] | NPV: 1.0<br>Sens: 1.0<br>Spec: 0.89<br>PPV: 0.70 | NPV: 0.99<br>Sens: 0.95<br>Spec: 0.94<br>PPV: 0.81 |
| Mass | 99 | 50.8% | 0.98 | 0.96 | 6.65% | 0.95 | 1.0 | 0.97 [0.95, 0.99] | NPV: 0.99<br>Sens: 0.97<br>Spec: 0.89<br>PPV: 0.79 | NPV: 0.98<br>Sens: 0.96<br>Spec: 0.94<br>PPV: 0.87 |
| Nodule | 130 | 47.0% | 0.96 | 0.95 | 7.73% | 0.96 | 1.0 | 0.94 [0.91, 0.97] | NPV: 0.91<br>Sens: 0.86<br>Spec: 0.89<br>PPV: 0.81 | NPV: 0.90<br>Sens: 0.82<br>Spec: 0.94<br>PPV: 0.88 |
| Pneumonia | 2 | 70.1% | 1.0 | 1.0 | 0.43% | 0.03 | 1.0 | 0.97 [0.93, 1.0] | NPV: 1.0<br>Sens: 1.0<br>Spec: 0.89<br>PPV: 0.07 | NPV: 1.0<br>Sens: 1.0<br>Spec: 0.94<br>PPV: 0.13 |
| Pneumothorax | 136 | 44.8% | 0.99 | 0.99 | 7.34% | 0.96 | 1.0 | 0.98 [0.97, 0.99] | NPV: 1.0<br>Sens: 0.99<br>Spec: 0.89<br>PPV: 0.84 | NPV: 0.96<br>Sens: 0.93<br>Spec: 0.94<br>PPV: 0.90 |
| Consolidation | 83 | 52.1% | 1.0 | 1.0 | 7.62% | 0.96 | 1.0 | 0.99 [0.98, 1.0] | NPV: 1.0<br>Sens: 1.0<br>Spec: 0.89<br>PPV: 0.76 | NPV: 0.99<br>Sens: 0.96<br>Spec: 0.94<br>PPV: 0.85 |
| Edema | 23 | 64.3% | 1.0 | 1.0 | 7.84% | 0.5 | 1.0 | 0.98 [0.95, 0.99] | NPV: 1.0<br>Sens: 1.0<br>Spec: 0.89<br>PPV: 0.47 | NPV: 1.0<br>Sens: 0.96<br>Spec: 0.94<br>PPV: 0.61 |
| Emphysema | 7 | 69.0% | 0.99 | 0.86 | 0.84% | 0.5 | 1.0 | 0.94 [0.83, 1.0] | NPV: 1.0<br>Sens: 1.0<br>Spec: 0.89<br>PPV: 0.21 | NPV: 0.99<br>Sens: 0.71<br>Spec: 0.94<br>PPV: 0.26 |
| Fibrosis | 13 | 66,9% | 1.0 | 1.0 | 1.63% | 0.75 | 1.0 | 0.99 [0.97, 1.0] | NPV: 1.0<br>Sens: 1.0<br>Spec: 0.89<br>PPV: 0.33 | NPV: 1.0<br>Sens: 0.92<br>Spec: 0.94<br>PPV: 0.46 |
| Pleural Thickening | 49 | 58.4% | 1.0 | 1.0 | 3.20% | 0.89 | 1.0 | 0.99 [0.98, 1.0] | NPV: 1.0<br>Sens: 0.98<br>Spec: 0.89<br>PPV: 0.65 | NPV: 0.98<br>Sens: 0.90<br>Spec: 0.94<br>PPV: 0.76 |
| Hernia | 5 | 69.6% | 0.99 | 0.8 | 0.42% | 0.05 | 1.0 | 0.89 [0.72, 0.99] | NPV: 1.0<br>Sens: 0.8<br>Spec: 0.89<br>PPV: 0.13 | NPV: 1.0<br>Sens: 0.8<br>Spec: 0.94<br>PPV: 0.22 |
| Other | 58 | 61.0% | 0.93 | 0.78 | 1.38% | 0.75 | 1.0 | 0.78 [0.70, 0.85] | NPV: 0.87<br>Sens: 0.47<br>Spec: 0.89<br>PPV: 0.51 | NPV: 0.88<br>Sens: 0.50<br>Spec: 0.94<br>PPV: 0.67 |

**Supplementary Table 6. Comparison between publicly available labels and majority vote labels by 3 radiologists on the CXR-14 test set (810 images).**

| Public NLP labels <br><br> Majority vote of 3 radiologists | Atelectasis | Cardiomegaly | Effusion | Infiltration | Mass | Nodule | Pneumonia | Pneumothorax | Consolidation | Edema | Emphysema | Fibrosis | Pleural Thickening | Hernia | Other* | No Finding |
|---|---|---|---|---|---|---|---|---|---|---|---|---|---|---|---|---|
| Atelectasis | 68 | 9 | 77 | 46 | 27 | 26 | 2 | 79 | 21 | 2 | 24 | 7 | 24 | 1 | 0 | 73 |
| Cardiomegaly | 17 | 22 | 26 | 15 | 4 | 3 | 1 | 6 | 7 | 1 | 2 | 2 | 7 | 0 | 0 | 17 |
| Effusion | 42 | 5 | 85 | 33 | 24 | 20 | 3 | 47 | 18 | 1 | 10 | 5 | 23 | 1 | 0 | 55 |
| Infiltration | 8 | 1 | 13 | 13 | 8 | 6 | 1 | 16 | 6 | 1 | 6 | 1 | 2 | 0 | 0 | 11 |
| Mass | 9 | 0 | 15 | 14 | 33 | 30 | 2 | 19 | 5 | 2 | 4 | 1 | 4 | 2 | 0 | 24 |
| Nodule | 12 | 2 | 23 | 16 | 23 | 38 | 0 | 18 | 6 | 2 | 8 | 6 | 6 | 1 | 0 | 42 |
| Pneumonia | 0 | 0 | 0 | 1 | 0 | 0 | 0 | 0 | 0 | 0 | 0 | 0 | 0 | 0 | 0 | 1 |
| Pneumothorax | 19 | 0 | 21 | 10 | 15 | 13 | 1 | 76 | 3 | 1 | 23 | 1 | 5 | 0 | 0 | 29 |
| Consolidation | 15 | 2 | 28 | 16 | 11 | 7 | 2 | 10 | 12 | 0 | 2 | 0 | 6 | 0 | 0 | 24 |
| Edema | 3 | 6 | 3 | 10 | 2 | 2 | 1 | 1 | 2 | 0 | 1 | 1 | 2 | 0 | 0 | 3 |
| Emphysema | 0 | 0 | 1 | 1 | 0 | 1 | 0 | 1 | 0 | 0 | 1 | 0 | 0 | 0 | 0 | 5 |
| Fibrosis | 2 | 0 | 1 | 4 | 0 | 1 | 0 | 4 | 1 | 0 | 0 | 1 | 1 | 0 | 0 | 4 |
| Pleural Thickening | 7 | 2 | 17 | 5 | 4 | 6 | 0 | 11 | 3 | 0 | 6 | 3 | 10 | 1 | 0 | 9 |
| Hernia | 2 | 0 | 1 | 1 | 0 | 0 | 0 | 0 | 0 | 0 | 0 | 0 | 1 | 2 | 0 | 2 |
| *Other | 6 | 3 | 4 | 8 | 5 | 3 | 1 | 6 | 3 | 0 | 1 | 3 | 4 | 0 | 0 | 29 |
| No Finding | 11 | 7 | 9 | 16 | 2 | 2 | 4 | 2 | 4 | 2 | 1 | 6 | 5 | 1 | 0 | 179 |

*Note, "Other" was not part of the public labels, and one that we added to indicate findings not covered by CXR-14's original 14 conditions, and for CXRs where the radiologists did not have a majority opinion regarding the specific finding.

**Supplementary Table 7. Quantitative evaluation of two workflows: (A) sequential DLS-Radiologist, and (B) comparison of DLS and sequential DLS-Radiologist in distinguishing normal and abnormal CXRs across six datasets.** A, The performance of radiologist reviewing cases after DLS' selection of abnormal CXRs for prioritized review across 6 datasets. B, Comparison of DLS and sequential DLS-Radiologist with non-inferiority test and percentages of potential caseload reduction.

**A**

| Scenario | Dataset | Performance of Combined Radiologist + DLS (with high-sensitivity operating point) | | | | | |
|---|---|---|---|---|---|---|---|
| | | No. predicted negative (%) | NPV | Sensitivity | No. predicted positive (%) | PPV | Specificity |
| Abnormality detection | DS-1 | 6,622 (85.5%) | 0.85 (0.85-0.86) | 0.48 (0.46-0.50) | 1,125 (14.5%) | 0.79 (0.76-0.81) | 0.96 (0.95-0.96) |
| | | 6,437 (83.1%) | 0.87 (0.86-0.88) | 0.54 (0.52-0.56) | 1,310 (16.9%) | 0.76 (0.74-0.78) | 0.95 (0.94-0.95) |
| | CXR-14 | 297 (36.7%) | 0.71 (0.66-0.76) | 0.85 (0.82-0.88) | 513 (63.3%) | 0.96 (0.94-0.98) | 0.91 (0.88-0.95) |
| | | 334 (41.2%) | 0.66 (0.61-0.71) | 0.80 (0.77-0.83) | 476 (58.8%) | 0.98 (0.96-0.99) | 0.95 (0.92-0.98) |
| Unseen disease: TB | TB-1 | 295 (63.9%) | 0.73 (0.68-0.78) | 0.67 (0.62-0.74) | 167 (36.1%) | 0.97 (0.94-0.99) | 0.98 (0.96-1.0) |
| | TB-2 | 90 (67.7%) | 0.88 (0.81-0.94) | 0.79 (0.68-0.90) | 43 (32.3%) | 0.98 (0.92-1.0) | 0.99 (0.96-1.0) |
| Unseen disease: COVID-19 | COV-1 | 1,196 (65.8%) | 0.78 (0.76-0.80) | 0.55 (0.51-0.59) | 623 (34.2%) | 0.51 (0.47-0.55) | 0.75 (0.73-0.78) |
| | COV-2 | 353 (58.3%) | 0.61 (0.57-0.66) | 0.53 (0.47-0.58) | 252 (41.7%) | 0.60 (0.55-0.66) | 0.68 (0.64-0.74) |

**B**

| Scenario | Dataset | Number Positives | Radiologist Sensitivity ("A") (95% CI) | Radiologist + DLS Sensitivity ("B") (95% CI) | Delta ("A"-"B") (95%CI) | Non-inferiority p-value | % caseload reduction |
|---|---|---|---|---|---|---|---|
| Abnormality detection | DS-1 | 1845 | 0.48 (0.46-0.51) | 0.48 (0.46-0.50) | 0.005 (0.002, 0.009) | **<0.00001** | 2,313 (29.9%) |
| | | | 0.54 (0.52-0.57) | 0.54 (0.52-0.56) | 0.004 (0.0001, 0.008) | **<0.00001** | |
| | CXR-14 | 578 | 0.87 (0.84-0.89) | 0.85 (0.82-0.88) | 0.01 (0.002, 0.02) | **<0.00001** | 194 (24.0%) |
| | | | 0.81 (0.78-0.84) | 0.80 (0.77-0.83) | 0.01 (0.0008, 0.02) | **<0.00001** | |
| Unseen disease: TB | TB-1 | 241 | 0.70 (0.65-0.76) | 0.67 (0.62-0.74) | 0.02 (0.002, 0.05) | 0.0062 | 199 (43.1%) |
| | TB-2 | 53 | 0.79 (0.68-0.90) | 0.79 (0.68-0.90) | 0.0 (-0.05, 0.05) | **<0.00001** | 51 (38.3%) |
| Unseen disease: COVID-19 | COV-1 | 580 | 0.55 (0.51-0.59) | 0.55 (0.51-0.59) | 0.0 (-0.005, 0.005) | **<0.00001** | 109 (5.9%) |
| | COV-2 | 288 | 0.53 (0.48-0.59) | 0.53 (0.47-0.58) | 0.003 (-0.008, 0.02) | **<0.00001** | 59 (9.8%) |